\documentclass[aps,prd,reprint,groupedaddress,
amsmath,amssymb]{revtex4-2}

\usepackage{graphicx}
\usepackage[table]{xcolor}
\usepackage{color,soul}
\usepackage{siunitx}

\newcommand{\dg}{\Delta g}

\definecolor{cardinal}{rgb}{0.77, 0.12, 0.23}
\definecolor{lightblue}{RGB}{0,114,189}
\definecolor{lightgray}{gray}{0.85}
\definecolor{blue}{rgb}{0.36, 0.54, 0.66}
\definecolor{amaranth}{rgb}{0.9, 0.17, 0.31}
\definecolor{pink}{rgb}{0.57, 0.36, 0.51}
\definecolor{maroon}{rgb}{0.76, 0.13, 0.28}
\definecolor{frenchlila}{rgb}{0.53, 0.38, 0.56}

\usepackage[pagebackref=false,
			hyperindex=true,
			citecolor=cardinal,
			colorlinks=true,
			linkcolor=blue,
			urlcolor=blue]{hyperref}

\begin{document}
	

\title{Detection and characterization of instrumental transients in LISA Pathfinder and their projection to LISA}



\author{Quentin Baghi$^1$,$^2$}
\author{Natalia Korsakova$^3$,$^4$}
\author{Jacob Slutsky$^1$}
\author{Eleonora Castelli$^5$}
\author{Nikolaos Karnesis$^6$}
	\author{Jean-Baptiste Bayle$^7$}

\affiliation{$^1$NASA Goddard Space Flight Center, 8800 Greenbelt Rd, Maryland, USA}
\affiliation{$^2$IRFU, CEA, Universit\'{e} Paris-Saclay, F-91191 Gif-sur-Yvette, France}
\affiliation{$^3$ARTEMIS, Observatoire de la Côte d’Azur, Bd de l'Observatoire, BP 4229, 06304 Nice, France}
\affiliation{$^4$SYRTE, Observatoire de Paris, Université PSL, CNRS, Sorbonne Université, LNE, 75014 Paris, France}
\affiliation{$^5$Department of Physics, University of Trento, Via Sommarive 14, 38123 Povo, Italy}
\affiliation{$^6$Department of Physics, Aristotle University of Thessaloniki, Thessaloniki 54124, Greece}
\affiliation{$^7$Jet Propulsion Laboratory, California Institute of Technology, Pasadena CA 91109}

%


\date{\today}

\begin{abstract}
	The LISA Pathfinder (LPF) mission succeeded outstandingly in demonstrating key technological aspects of future space-borne gravitational-wave detectors, such as the Laser Interferometer Space Antenna (LISA). Specifically, LPF demonstrated with unprecedented sensitivity the measurement of the relative acceleration of two free-falling cubic test masses. Although most disruptive non-gravitational forces have been identified and their effects mitigated through a series of calibration processes, some faint transient signals of yet unexplained origin remain in the measurements. If they appear in the LISA data, these perturbations (also called glitches) could skew the characterization of gravitational-wave sources or even be confused with gravitational-wave bursts. For the first time, we provide a comprehensive census of LPF transient events. Our analysis is based on a phenomenological shapelet model allowing us to derive simple statistics about the physical features of the glitch population. We then implement a generator of synthetic glitches designed to be used for subsequent LISA studies, and perform a preliminary evaluation of the effect of the glitches on future LISA data analyses.
\end{abstract}
	
	
\maketitle

\section{\label{sec:intro}Introduction}
	
\begin{figure}[ht]
	\begin{center}
		\includegraphics[width=1.0\columnwidth, trim={0.5cm, 0.5cm, 1.6cm, 0.5cm}, clip]{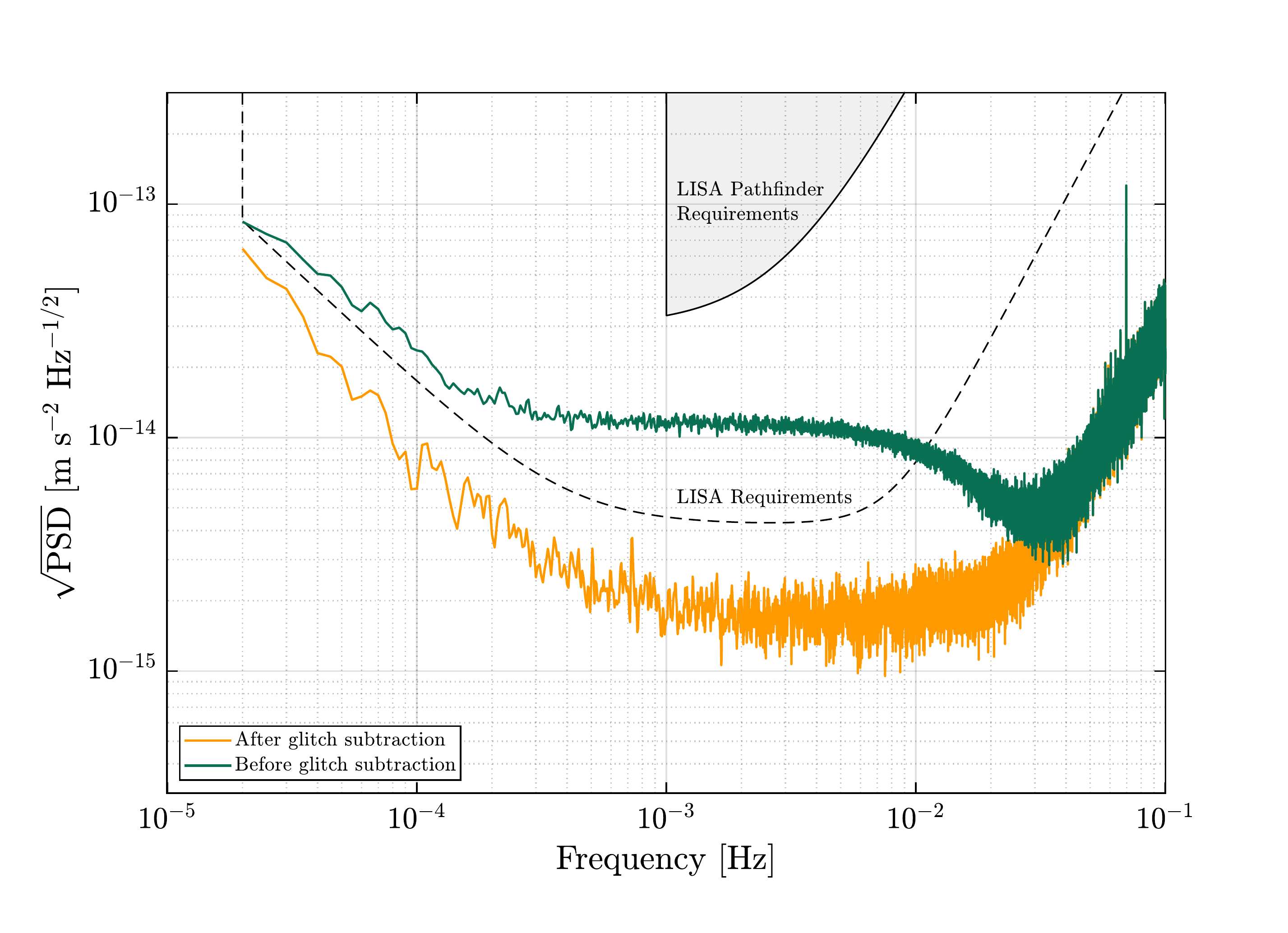}
		\caption{Amplitude spectral density ($\sqrt\mathrm{PSD}$) of parasitic differential acceleration of LPF TMs as a function of frequency for the best performance noise run, with (green) and without (orange) spurious signals.}
		\label{fig:deltag_psd}
	\end{center}
\end{figure}

LISA Pathfinder (LPF) was launched on December $\rm 3^{rd}$, 2015 as a technological demonstrator for the Laser Interferometer Space Antenna (LISA), a future space-based gravitational-wave observatory~\cite{lisamission}. The LPF mission measured the relative acceleration between two cubic free-falling test masses (TMs) housed within the spacecraft (SC), up to a precision of $30~\mathrm{fm~s^{-2} ~Hz^{-1/2}}$ for mHz frequencies, as depicted by the orange spectrum in Fig.~\ref{fig:deltag_psd}.
	
LPF demonstrated the feasibility of LISA free-fall precision requirement via the assessment of parasitic differential acceleration $\dg (t)$ between the two free-floating TMs, which constitutes the main science measurement of the LPF mission. This performance reached the required free-fall levels for LISA~\cite{lpf_prl, lpf_prl2}, and was far more sensitive than LPF's precision target. Several incremental corrections and calibrations were necessary to reach this better-than-expected goal~\cite{lpfcalibration}, along with the characterization of various subsystems. The satellite was successfully operated until its planned decommissioning on July $\rm 18^{th}$, 2017.
	
The TMs differential acceleration data $\dg (t)$, once cleaned of the inertial effects, were found to be affected by spurious non-gravitational forces of instrumental origin, which degraded LPF noise performance and limited its sensitivity. These perturbations, often referred to as \emph{glitches}, were transient in nature. To produce the results in Ref.~\citep{lpf_prl2}, they were subtracted from $\dg (t)$ data in order to recover the full LISA-like sensitivity, allowing to go from the green to the orange periodogram in Fig.~\ref{fig:deltag_psd}. The removal of these transients was particularly critical to obtain the full sensitivity at the lowest frequencies. Should they arise in LISA data, their presence could impact the characterization of gravitational-wave sources. Adequately accounting for them in the data modeling is therefore crucial both for noise and signal estimation.

The physical nature of these spurious events is still unknown. They manifested as uncorrelated non-periodic events during ordinary runs, occurring with varying intensity and duration. 
The glitch phenomenology presented here appears to be connected to a series of physical counterparts that are currently under investigation and are outside the aim of this work.
Hypotheses regarding their physical origin are under analysis and require further simulation and eventually on-ground testing at torsion pendulum facilities. Note that these unmodeled perturbations in the differential acceleration measurements are different from micrometeorids impacts considered by the authors of Ref.~\cite{Thorpe2019}, which are detected with satellite position telemetry. 

Despite the lack of a physical understanding of their origin, glitches can be modeled mathematically. In this work, we adopt an approach to allow for an automated and systematic glitch detection. We use generic basis functions called \textit{shapelets}~\cite{Berge2019} to form a template bank and detect events through matched filtering. Here, we only consider TM differential acceleration measurements, and we detect events without any assumption on their physical nature. The objective is two-fold: i) provide a mathematical and comprehensive study of glitch population that can be informative for LISA and ii) obtain a distribution of event parameters serving as an input for LISA simulation studies, as well as data challenges~\cite{LDC}. 
	
The article is outlined as follows. In Section~\ref{sec:data}, we start by describing the measurements used in this analysis. We then introduce the shapelet model and the detection method in Section~\ref{sec:method}. In Section~\ref{sec:search} we present the results of the glitch search and evaluate its performance, before analyzing the statistics of glitch parameters in Section~\ref{sec:statistics}. We show in Section~\ref{sec:pe} that these parameters can also be estimated through Bayesian analysis from matched filtering outputs, for an optimal modeling and removal. Finally, based on the statistics we obtain, we describe the design of a glitch generator readily usable for LISA studies in Section~\ref{sec:glitch_generator}.
	
\section{\label{sec:data}Analyzed measurements}

\subsection{\label{sec:datadesc}Description of the data}

The differential acceleration $\dg (t)$ is derived from the most sensitive measurement of the TMs position, performed via the differential interferometer $ o_{12}$~\cite{oms1,oms3}, as
\begin{equation}
	\label{eq:deltagfull}
	\dg (t) \equiv  \ddot{o}_{12} (t) - g_c(t) + \omega_2^2  o_{12} (t) + \Delta\omega^2_{12} o_1 (t),
\end{equation}
where the $\omega_i^2$ terms are the spring-like constants of noisy forces acting on the TMs, due to electrostatic, magnetic, and local gravity effects, while $g_c(t)$ are the commanded forces, known up to a multiplicative amplitude calibration constant.
	
Equation \eqref{eq:deltagfull} describes the dynamics along the $x$-axis, but does not include any inertial effects caused by cross-coupling with other degrees-of-freedom. Thus the understanding and accurate calibration of $\dg (t)$ required a series of dedicated experiments, designed to identify the dynamical relations between the three-body system constituted by the TMs and SC, and the various environmental noises contributing to the overall noise budget~\cite{lpfcalibration, wanner2017}.
	
After subtraction of the inertial effects, $\dg(t)$ includes the sensitive $x$-axis dynamics of LPF TMs free of the inertial forces acting on the system, allowing for evaluation of the LPF noise performance on \emph{noise-only} measurements. Noise-only measurements were acquired during mission operations, in-between the different calibration and system identification experiments, lasting up to two weeks at a time, in order to benchmark LPF performance.
	
The analysis in this paper only examines the noise-only $\dg$ measurements containing spurious signals, which are publicly available on the LPF legacy data archive~\cite{LPFLA}. We select continuous time series segments where no signal injections took place, and we use the developed detection method to identify the transient signals in the noise (see Section~\ref{sec:method} below). The actual time segments are listed in Appendix~\ref{apx:list_segments}.
	
\subsection{\label{sec:problem}Target instrumental perturbations}
	
The data exhibit spurious signals of varying duration and amplitudes. These artefacts are distinguishable from stationary Gaussian noise, and can be described by a transient process with short time duration compared to the overall measurement.
	
\begin{figure}[!ht]
	\centering
	\begin{tabular}{c}
	     \includegraphics[width=0.95\columnwidth]{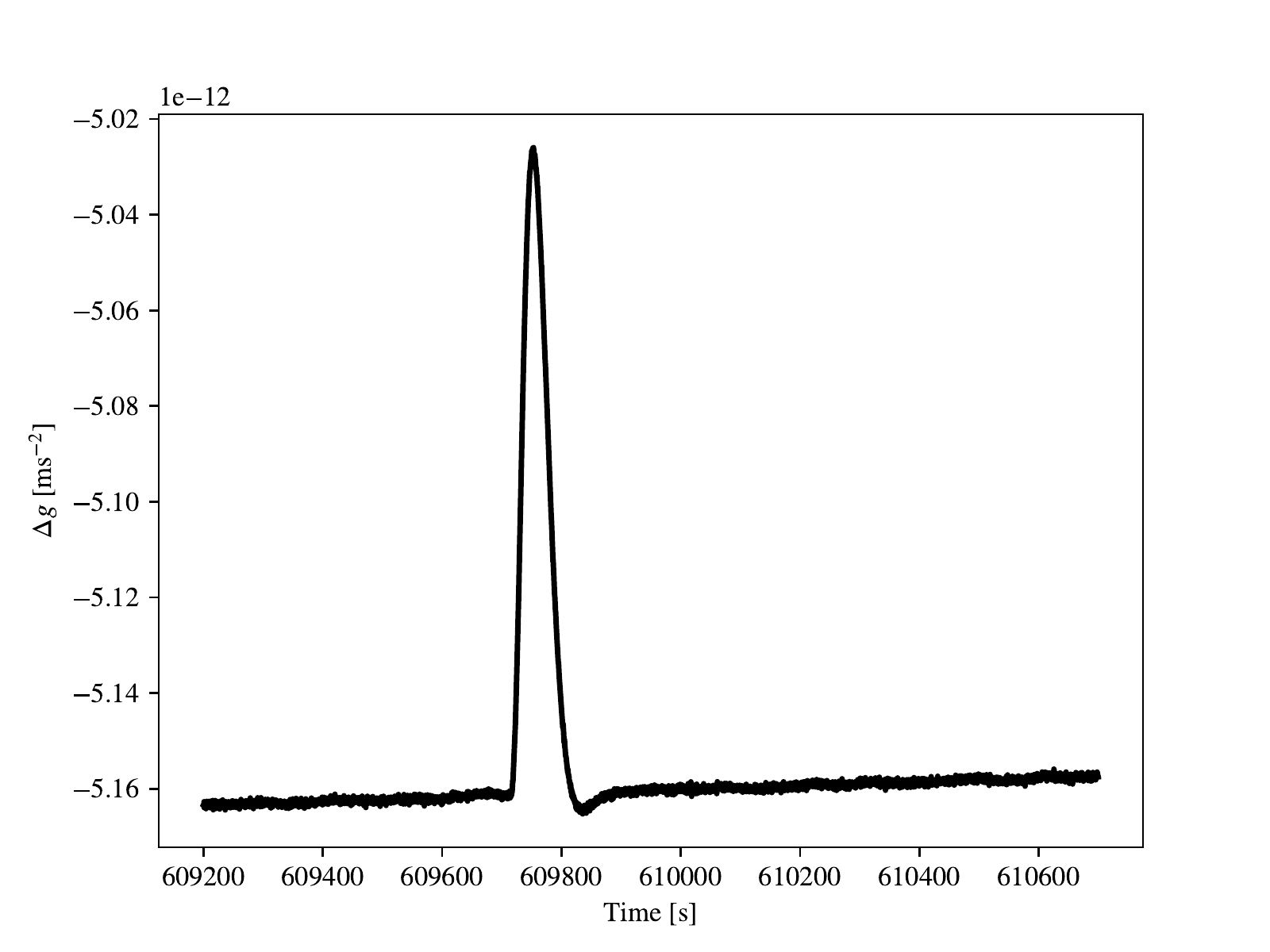}\\
	     (a) FRED glitch \label{fig:exp} \\
	     \includegraphics[width=0.95\columnwidth]{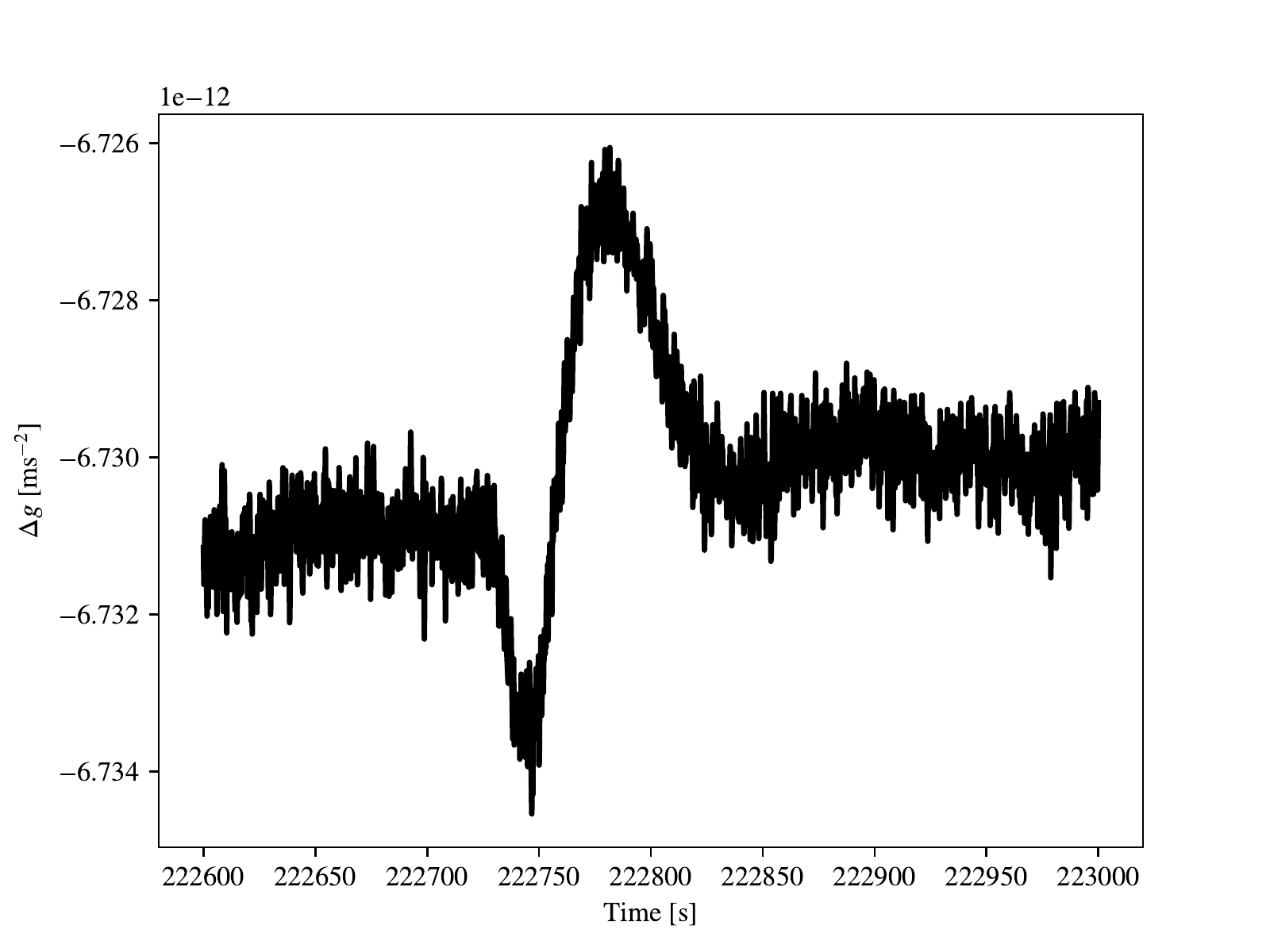} \\
	     (b) Sine-Gaussian glitch \label{fig:sG}
	\end{tabular}
	\caption{Example of two glitch types observed in LPF: fast rise and slow exponential decay type (upper panel) and singe-Gaussian type (bottom panel). Data have been low-pass filtered with a second-order Butterworth filter with a cut-off frequency of $10^{-2}$ Hz.}%
	\label{fig:twotypes}%
\end{figure}

We observe two types of glitches. The majority include a sharp rise followed by a slow exponential decay back to the previous acceleration trend. The initial slope can be a positive or negative acceleration. Adopting the gamma-ray burst terminology~\cite{Fishman1994}, we will refer to these glitches as fast rise and an exponential decay (FRED) events. An example is shown in Fig.~\ref{fig:exp}. The other common type of glitches, which occur less frequently, have a sine-Gaussian shape characterized by two sequential excursions with opposite signs, as shown in the example of Fig.~\ref{fig:sG}). Although this is beyond the scope of this work, ongoing analysis suggest that the two glitch types are connected to different physical phenomena \cite{Castelli2020, lpf_noiseperf3}.

During nominal mission operations, glitches appeared stochastically with a characteristic time interval. However, during the course of LPF mission, the system was cooled down in order to further investigate the effects contributing to the overall noise budget. Unexpectedly, this increased the glitch event rate substantially. We will henceforth refer to these measurements as \textit{cold runs}, and we will separately investigate them in addition to the ordinary noise runs.
	
\section{\label{sec:method}Detection method}

\subsection{\label{sec:model}Penalized likelihood model}

In order to make an inventory of transient signals in LPF $\Delta g$ data, we need a flexible framework to detect the corresponding excess of power. This can be done by decomposing glitch signals into a dictionary of functions where they are sparse, i.e. only a reduced number of non-zero coefficients is necessary to accurately represent the signal.
	
The general model of the measured data can be written as follows:
\begin{equation}
	\mathbf{y} = \mathbf{s} + \mathbf{n},
\end{equation}
where $\mathbf{y}$ is the observed time series of size $N$ sampled at frequency $f_s$, $\mathbf{s}$ is the glitch signal, and $\mathbf{n}$ is the realization of a stochastic random variable assumed to follow a zero-mean, stationary Gaussian distribution.
	
Formally, we design the detection pipeline by finding the maximum of a penalized likelihood function whose logarithm can be written as
\begin{equation}
	\label{eq:penalized_log_likelihood}
	L\left(\mathbf{\theta}\right) = \log p\left(\mathbf{y} | \mathbf{\theta} \right) - \mathrm{pen}_{\lambda}\left(\mathbf{\theta}\right),
\end{equation}
where $p_{\theta}(\mathbf{y} | \mathbf{\theta})$ is the unpenalized likelihood function of the model parameters $\mathbf{\theta}$, equal to the Gaussian probability density function in this problem. The second term $\mathrm{pen}_{\lambda}(\theta)$ is the penalty function, which controls the thresholding of the function amplitudes. The vector $\mathbf{\theta}$ is the vector of parameters to be estimated, and $\lambda$ is the regularization parameter, controlling the model quality in representing the data and the sparsity level.

For a Gaussian zero-mean distribution, the unpenalized likelihood writes
\begin{equation}
	\label{eq:gaussian_likelihood}
	p_{\theta}(\mathbf{y}) =  \frac{\exp\left\{ - \frac{1}{2}\left( \mathbf{y} -  \mathbf{s}(\theta_s) \right)^{\dag} \mathbf{\Sigma}\left(\mathbf{\theta}_n\right)^{-1}\left( \mathbf{y} - \mathbf{s}(\theta_s)  \right) \right\}}{\sqrt{ (2\pi)^{N} \left| \mathbf{\Sigma}\left(\mathbf{\theta}_n\right) \right|}},
\end{equation}
where $\mathbf{\Sigma}$ is the covariance matrix of the noise vector $\mathbf{n}$, $\mathbf{x}^{\dag}$ refers to the Hermitian conjugate of any vector or matrix $\mathbf{x}$ and $\left| \mathbf{X}\right|$ is the determinant of any matrix $\mathbf{X}$. Note that in this formulation, the model parameter vector includes both signal parameters $\mathbf{\theta}_s$ and noise parameters $\mathbf{\theta}_{n}$, and can therefore be written as $\mathbf{\theta} \equiv \left( \mathbf{\theta}_s, \mathbf{\theta}_{n}\right)$.
We further assume that the noise is wide-sense stationary, with power spectral density (PSD) $S_{n}(\mathbf{\theta}_{n}, f)$, so that its covariance can be approximated by a circulant matrix that is diagonalizable in the Fourier basis:
\begin{equation}
	\mathbf{\Sigma} = \mathbf{F}^{\dag} \mathbf{\Lambda} \mathbf{F},
\end{equation}
where $\mathbf{F}$ is the discrete Fourier transform matrix with elements $F_{kn} = e^{-2 j \pi k n / N}$ and $\mathbf{\Lambda}$ is a diagonal matrix whose diagonal elements are proportional to the PSD $\Lambda_{kk} = N f_{s} S_{n}(\mathbf{\theta}_{n}, f_{k})$

In the following, we choose a penalty function equal to the $l_{0}$ pseudo norm of the coefficient vector, which imposes the sparsest constraint on the representation, and is well adapted to sharp features:
\begin{equation}
	\label{eq:penalty}
	\mathrm{pen}_{\lambda}(\mathbf{\alpha}) = \lambda  \lVert \mathbf{\alpha} \rVert_{0} = \lambda \lim_{p \rightarrow 0} \sum_{i=0}^{N-1} \left| \alpha_{i} \right|^{p}.
\end{equation}
The $l_{0}$-norm of a vector gives its number of non-zero coefficients, thus minimizing it enforces sparsity.

\subsection{\label{sec:shapelets}Modeling transient signals using shapelets}

As we do not postulate any physical process underlying the glitch phenomena, we choose to adopt a representative phenomenological model. Given previous observations, this model should have the following characteristics: it should allow for i) a sharp rise, ii) an exponential decay and iii) one or a few oscillations. Most of wavelet functions, including Gaussian-shape sinusoids, are not well suited to that kind of transients because of their symmetric shape. Asymmetric functions defined for positive times only are preferable. In addition, the model should be general enough to describe the various types of glitches that have been cataloged. 

A good option that meets these requirements is the exponential shapelet function described in Ref.~\cite{Berge2019}. This is a collection of exponentially damped functions that are eigen-wavefunctions of the normalized 1D-hydrogen atom. They can be defined in the time domain as follows:
\begin{equation}
	\label{eq:shapelet_model}
	\psi_{n}(t) = c_{n} \frac{2t}{n} e^{-\frac{t}{n}} L^{1}_{n-1}\left(\frac{2t}{n} \right) h_{+}(t),
\end{equation}
where $c_{n} = (-1)^{n-1}n^{-\frac{3}{2}}$ is a normalizing factor ensuring that the quadratic sum of the components is equal to 1, and $h_{+}(t)$ is the Heaviside function which is equal to $1$ for $t \geq 0$, $0$ otherwise.
A glitch perturbation is then modeled by a finite linear combination of shapelets:
\begin{equation}
\label{eq:linear_combination}
g(t) = \sum_{i=0}^{P-1}  \alpha_{i}  \psi_{n}\left( \frac{t - \tau_{i}}{\beta_{i}}\right).
\end{equation}
The parameters to be estimated for each shapelet component are therefore the scale parameter $\beta$, which acts as a characteristic damping time and hence is related to the glitch duration, the arrival time $\tau$ and the amplitude $\alpha$. We denote as $\mathbf{\theta}_s \equiv \left( \beta, \, \mathbf{\tau}, \, \mathbf{\alpha} \right)$ the vector gathering the shapelet parameters.

\subsection{Matching pursuit algorithm}

A common approach for signal detection in gravitational-wave data analysis is to perform a matched filtering of the data with a template bank, correlating the observed signal with a parametric waveform model. This method is equivalent to calculating the ratio between two likelihoods: the likelihood in the numerator corresponds to the hypothesis that a signal is present, and the one in the denominator corresponds to the null hypothesis (absence of signal). 

However, when the signal is assumed to be the superimposition of several non-orthogonal waveform atoms, direct matched filtering can be cumbersome. Instead, here we use an iterative version called matching-pursuit \cite{Mallat1993}, where matched filtering is applied at each iteration to the residuals of the fit at the previous iteration. The final estimation is given by the sum of the signals estimated individually. It can be shown that this algorithm approximately solves the maximum penalized likelihood problem
\begin{equation}
	\mathbf{\theta}_{s, \mathrm{opt}} = \mathrm{argmax}_{\mathbf{\theta}_s} L\left(\mathbf{\theta}_s\right),
\end{equation}
where $L\left(\mathbf{\alpha}\right)$ is the penalized log-likelihood given in Eq.~(\ref{eq:penalized_log_likelihood}) with fixed covariance parameter $\mathbf{\theta}_{n}$ (assumed to be known) and with penalty function given by Eq.~(\ref{eq:penalty}).

Let us consider iteration $i$, and let us label $\mathbf{r}^{(i)}$ the residuals from the previous iteration. We assume that we already found $i-1$ components fitting the signal, and we now want to find the next significant element, if any, to add to the sum in Eq.~(\ref{eq:linear_combination}). We are looking for the value of $\mathbf{\theta}_s$ that best fits the signal left in $\mathbf{r}^{(i)}$.
This is done by matched filtering the residuals with the waveform model $h\left(t, \mathbf{\theta}\right)$ given by 
		\begin{equation}
	h(t, \tau, \beta) \equiv  \psi_{n}\left( \frac{t - \tau}{\beta} \right),
\end{equation}
i.e., finding the set of parameters $\tau, \beta$ that maximize the output signal-to-noise ratio (SNR)
\begin{equation}
	\label{eq:snr}
	\rho_{i} = \frac{\langle \mathbf{h} | \mathbf{r}^{(i)} \rangle }{\sqrt{\langle \mathbf{h} | \mathbf{h} \rangle }},
\end{equation}
where we defined the scalar products $\langle \cdot | \cdot \rangle$ as
\begin{equation}
	\label{eq:scalar_product}
	\langle \mathbf{h} | \mathbf{r}^{(i)} \rangle = \sum_{k=0}^{N-1} \frac{\tilde{h}^{*}(f_k, \tau, \beta, n) \tilde{r}_{k}^{(i)} }{S(f_k)}.
\end{equation}
The symbol $\mathbf{\tilde{x}}$ represents the discrete Fourier transform (DFT) of any vector $\mathbf{x}$, defined as
\begin{equation}
	\tilde{x}_{k} \equiv \frac{1}{\sqrt{N f_{s}}}\sum_{m=0}^{N-1} x_{m} e^{-2 j \pi \frac{mk}{N}},
\end{equation}
where the normalization depends on the time series size $N$ and the sampling frequency $f_s$, and is chosen such that the square modulus of $\tilde{x}_{k}$ is homogeneous to a power spectral density.
If we denote $\tilde{\psi}_{n}(f)$ the continuous Fourier transform of the shapelet function in Eq.~(\ref{eq:shapelet_model}), we can write
\begin{equation}
\label{eq:frequency_template}
\tilde{h}(f, \tau, \beta, n) \approx \beta \tilde{\psi}_{n}\left( \beta f \right) e^{-2 j \pi f \tau}.
\end{equation}
We now fix the shapelet order to $n = 1$. This choice greatly simplifies the model and hereby lightens the computational cost of matched filtering. The counterpart is that we may need more shapelet components to fit the rare events with multiple oscillations. 

We approximate the DFT of the shapelet function by its continous Fourier transform evaluated at discrete Fourier frequencies $f_k$, as given in Ref.~\cite{Berge2019}:
\begin{equation}
	\tilde{\psi}_{n}(f_k) = (-1)^{n} \sqrt{\frac{2n}{\pi}} \frac{\left(n k - j \right)^{2n}}{\left( (nk)^{2} + 1\right)^{n+1}}.
\end{equation}
Plugging Eq.~(\ref{eq:frequency_template}) into Eq.~(\ref{eq:scalar_product}) yields
\begin{equation}
\label{eq:scalar_product2}
\langle \mathbf{h} | \mathbf{r}^{(i)} \rangle = \sum_{k=0}^{N-1} \frac{\beta \tilde{\psi}^{*}_{n}\left( \beta f_{k} \right)  \tilde{r}_{k}^{(i)} }{S(f_k)} e^{2 j \pi f_{k} \tau}.
\end{equation}
Note that we choose the grid of arrival times $\tau_{m} = m / f_s $ to be the same as the time series sampling cadence, then Eq.~(\ref{eq:scalar_product2}) can be efficiently computed with an inverse Fast Fourier transform (FFT) algorithm.

\subsection{Detection and false alarm probability}

Here we derive the detection threshold used in the matched filtering process. In gravitational-wave data analysis, a common way to do that is through the F-statistics, which is the likelihood in Eq.~(\ref{eq:gaussian_likelihood}) maximized with respect to the parameters on which the model depends linearly, called extrinsic parameters~\cite{Jaranowski2012}. In our problem, the extrinsic parameters are the amplitudes of the shapelet functions $\alpha_{i}$. Maximizing the logarithm of Eq.~(\ref{eq:gaussian_likelihood}) with respect to $\mathbf{\alpha}$ yields
\begin{equation}
	\label{eq:gaussian_likelihood_reduced}
	F_{\mathbf{y}}\left(\mathbf{\theta}_{\mathrm{intr}}\right) =  \frac{1}{2} \frac{\left| \langle \mathbf{h} | \mathbf{y} \rangle \right|^{2}}{\langle \mathbf{h} | \mathbf{h}\rangle},
\end{equation}
where $\mathbf{\theta}_{\mathrm{intr}} \equiv \left(\beta, \tau \right)$ denotes the intrinsic parameters, i.e. all other parameters different than the extrinsic ones.
The scalar product $ \langle \mathbf{h} | \mathbf{y} \rangle$ involves a weighted sum of the data DFT components $\tilde{y}_{k}$ which, in the absence of signal, follows a zero-mean Gaussian distribution. Therefore the scalar product follows a zero-mean Gaussian distribution with covariance $\langle \mathbf{h} | \mathbf{h}\rangle^{-1}$, and $2 F_{\mathbf{y}}$ is a chi-squared distribution with 2 degrees of freedom~\cite{Cutler2005}. The cumulative probability density of $F_{\mathbf{y}}$ is
\begin{equation}
	p\left(F \leq x \right) = \Gamma_\text{r}\left(1, x\right)
\end{equation}
where $\Gamma_\text{r}$ is the regularized gamma function, given by
\begin{equation}
	\Gamma_\text{r}\left(s, x\right) = \frac{1}{\Gamma(s)} \int _{0}^{x}t^{s-1} \,\mathrm {e}^{-t}\,{\rm {d}}t,
\end{equation}
where $\Gamma$ is the gamma function.
	
In practice, we will calculate the F-statistics for a given grid of values of the arrival time $\tau$ and the damping parameter $\beta$. Assuming that the different variables $F_{i} \equiv F(\tau_{i}, \beta_{i})$ are uncorrelated, the probability that neither of them will exceed the threshold $F_{0}$ is
\begin{eqnarray}
	p\left( \max_{i} F(\tau_{i}, \beta_{i}) \leq F_{0} \right) &=& \prod_{i=0}^{q} p\left( F(\tau_{i}, \beta_{i}) \leq F_{0} \right) \nonumber \\
	& = & \left[ \Gamma_\text{r}\left( 1, F_{0} \right) \right]^{q},
\end{eqnarray}
where $q$ is the number of explored parameters.
The total false alarm probability is therefore
\begin{equation}
	P_{F, \mathrm{tot}} = 1 - \left[ \Gamma_\text{r}\left( 1, F_{0} \right) \right]^{q}.
\end{equation}
If we impose a total false alarm probability $\alpha$, then the threshold must be such that
\begin{equation}
	F_{0}  = \Gamma_\text{r}^{-1}\left( 1, \left(1 - \alpha\right)^{\frac{1}{q}} \right).
\end{equation}
For large $q$, we can approximate the last equality by
\begin{equation}
	F_{0} = \log \left(\frac{q}{\alpha}\right).
\end{equation}
We impose to have a $\alpha = 0.01\%$ false alarm probability, meaning that there is $0.01\%$ chance that the detection could be triggered by Gaussian random noise only. The corresponding threshold for a 2.5-day segment sampled every 10 seconds with a grid of 80 values for $\beta$ is nearly $\mathrm{SNR} = \sqrt{F_{0}} \sim 5$. Adopting this value, we chose to focus on the robustness of the detection over its completeness, in particular to be conservative with respect to power spectral density estimation errors.

\section{\label{sec:search}Application to LPF noise runs}

The matching pursuit method described in the previous section is applied to 56 LPF $\Delta g$ noise-only measurements acquired between March 2016 and July 2017, lasting 1 to 10 days. Some of them contain one or several transient events. In this section, we describe the tuning of the search and its outputs. 

\subsection{Pre-processing}

Prior to running the algorithm, we divide each noise run in segments of $T_{\mathrm{seg}} = 2.5$ days (or smaller, when the noise measurement run itself is shorter). This segmentation allows us to bound the memory cost of the matched filtering, while encompassing the longest events in the search window. After computing the SNR as a function of the arrival time, we crop the SNR time series by $T_{\mathrm{seg}} / 8 = 7.5$ hours at the start and at the end, to avoid edge effects (see, for example, \cite{allen_findchirp_nodate}). As a result, there is an overlap of $2 / 8 = 25 \%$ between two consecutive segments. 

Afterwards, we perform a raw estimate of the noise PSD, by fitting a spline to the log-periodogram. This fit is made by taking the median of the PSDs estimated on 8 subsegments, improving the robustness against glitches. This provides a first estimate for the matching pursuit run, during which the PSD estimate will be updated using the residuals of the fit at each iteration. The parametrization of the search is described below.

\subsection{Parametrization of the search}
The parameter grid that we explore is evenly sampled in logarithmic space of $\beta$, with 80 values in the interval $[0.1,\, 50000]~\mathrm{s}$. The lower bound corresponds to the sampling time, and the upper bound corresponds to about twice the duration of the longest glitch. In addition, we evenly sample arrival times at the same rate as the input time series (0.1 second). The order of the shapelet model is fixed to $n=1$ to mitigate the computational cost of the search. Since a single shapelet of order 1 may not be sufficient to account for one glitch shape, we allow for several shapelet elements to describe a single event. For example, sine-Gaussian events are usually modelled with two elements of the basis.
Note that a new shapelet component is added to the model only if the associated SNR in the residuals of the previous iteration exceeds the detection threshold, thereby preventing overfitting.

The iterations of the matching pursuit algorithm stop when there are no more signal in the residuals that exceeds the SNR threshold. 
The estimated glitch signal is then the sum of all detected components, as given by Eq.~(\ref{eq:linear_combination}).

\subsection{Search results}

\begin{figure}[!ht]
	\centering
	\begin{tabular}{c}
	\includegraphics[width=1.0\columnwidth,  trim={0.2cm, 0.3cm, 0.1cm, 0.3cm}, clip]{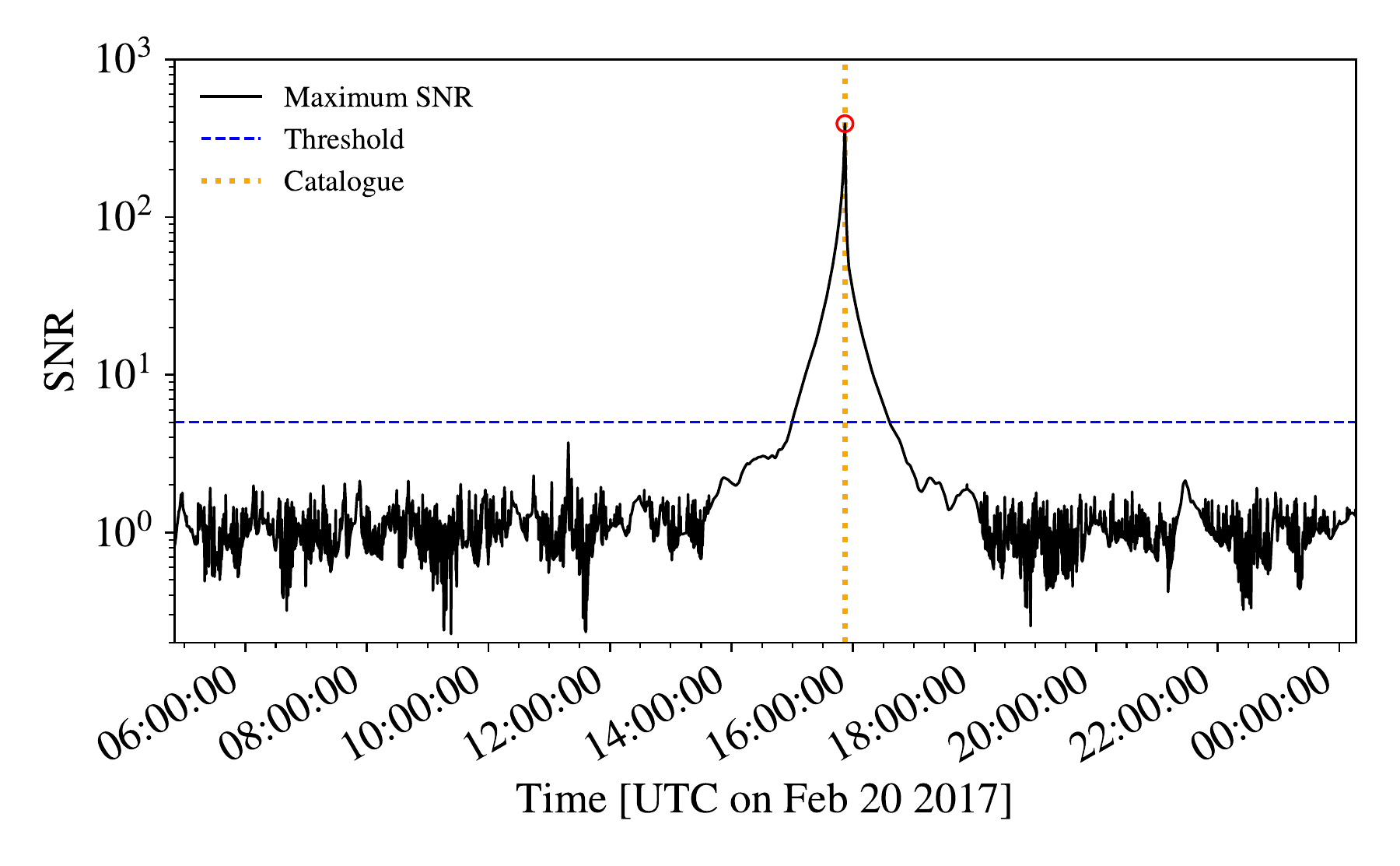} \\
	\includegraphics[width=1.0\columnwidth,  trim={0.2cm, 0.3cm, 0.1cm, 0.3cm}, clip]{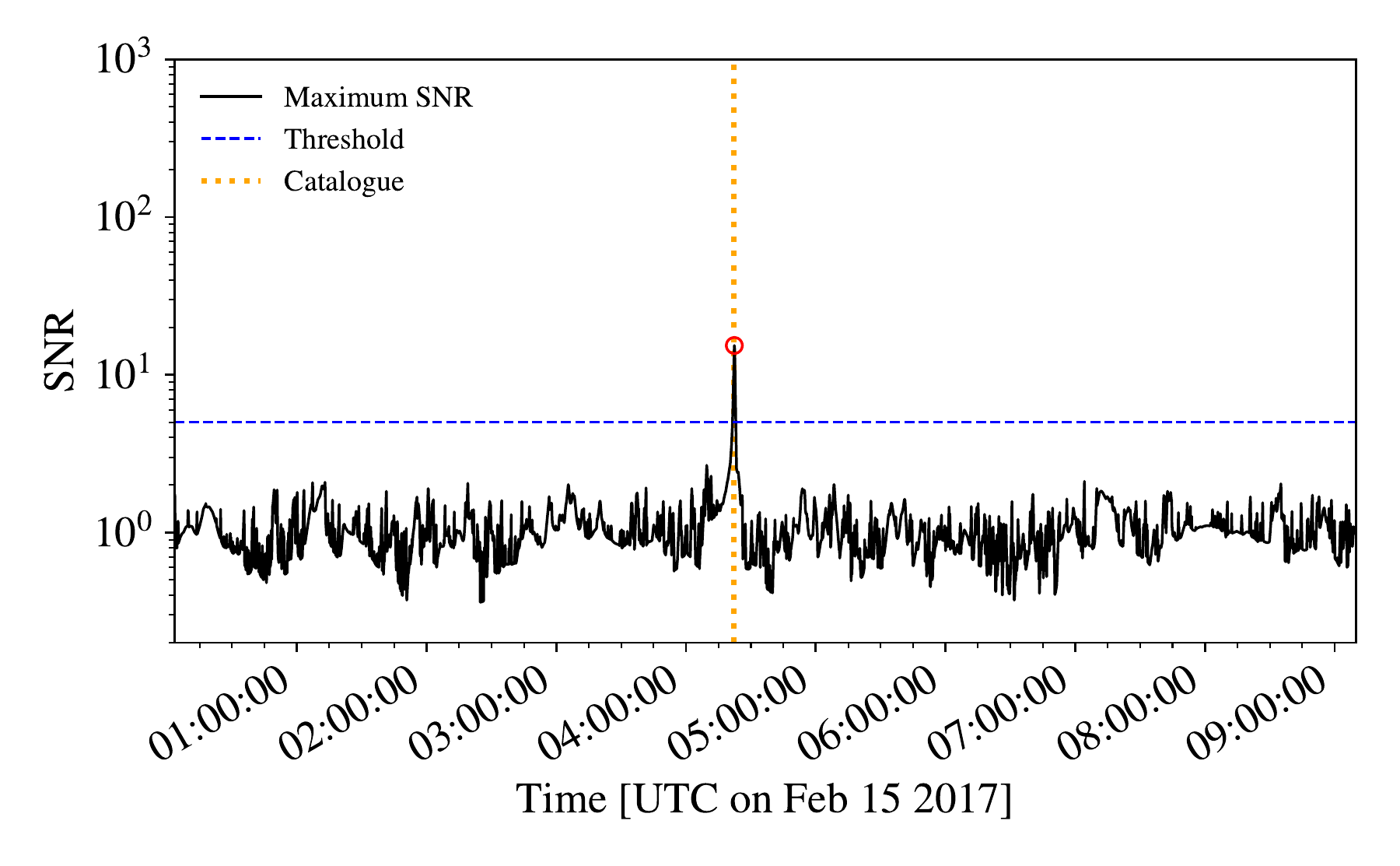}
	 \end{tabular}
	\caption{SNR time series maximized over $\beta$, obtained at the first iteration of the matching pursuit algorithm, for the third (upper panel) and first segment (lower panel) of the 66\textsuperscript{th} noise run. The solid black line is the computed SNR, and the horizontal blue dashed line corresponds to the detection threshold $\text{SNR} = 5$. Red circles localize detected events. Vertical dotted lines in orange corresponds to the arrival times that were initially recorded in the legacy catalog. }%
	\label{fig:snr_time_series_66}%
   
\end{figure}

At each iteration, the matching pursuit algorithm computes the SNR time series from the residuals of the previous iteration, using the current estimate of the noise PSD. In Fig.~\ref{fig:snr_time_series_66}, we present the SNR values, maximized over $\beta$, computed at the first iteration of the third search segments in the noise run covering the period between February 13\textsuperscript{th} 2017 and March 2\textsuperscript{nd} 2017 (or noise run number 66). We choose this excerpt because it corresponds to the best noise performance achieved~\cite{lpf_prl2}, and we zoom in the interval where two events of different amplitudes are clearly visible.

For these events, the matching pursuit detections coincide with events recorded in the legacy catalog made to produce the data in Ref.~\cite{lpf_prl2}. In this earlier analysis, glitch detection was performed visually using low-pass filtered data. Then, glitches were subtracted using ad hoc models based on exponential functions or filtered impulses~\cite{Castelli2020}. This comparison confirms our ability to automatically detect transients that were manually highlighted.

\begin{figure}[h!t]
	\centering
	\begin{tabular}{c}
		\includegraphics[width=1.0\columnwidth,  trim={0.0cm, 0.3cm, 0.1cm, 0.3cm}, clip]{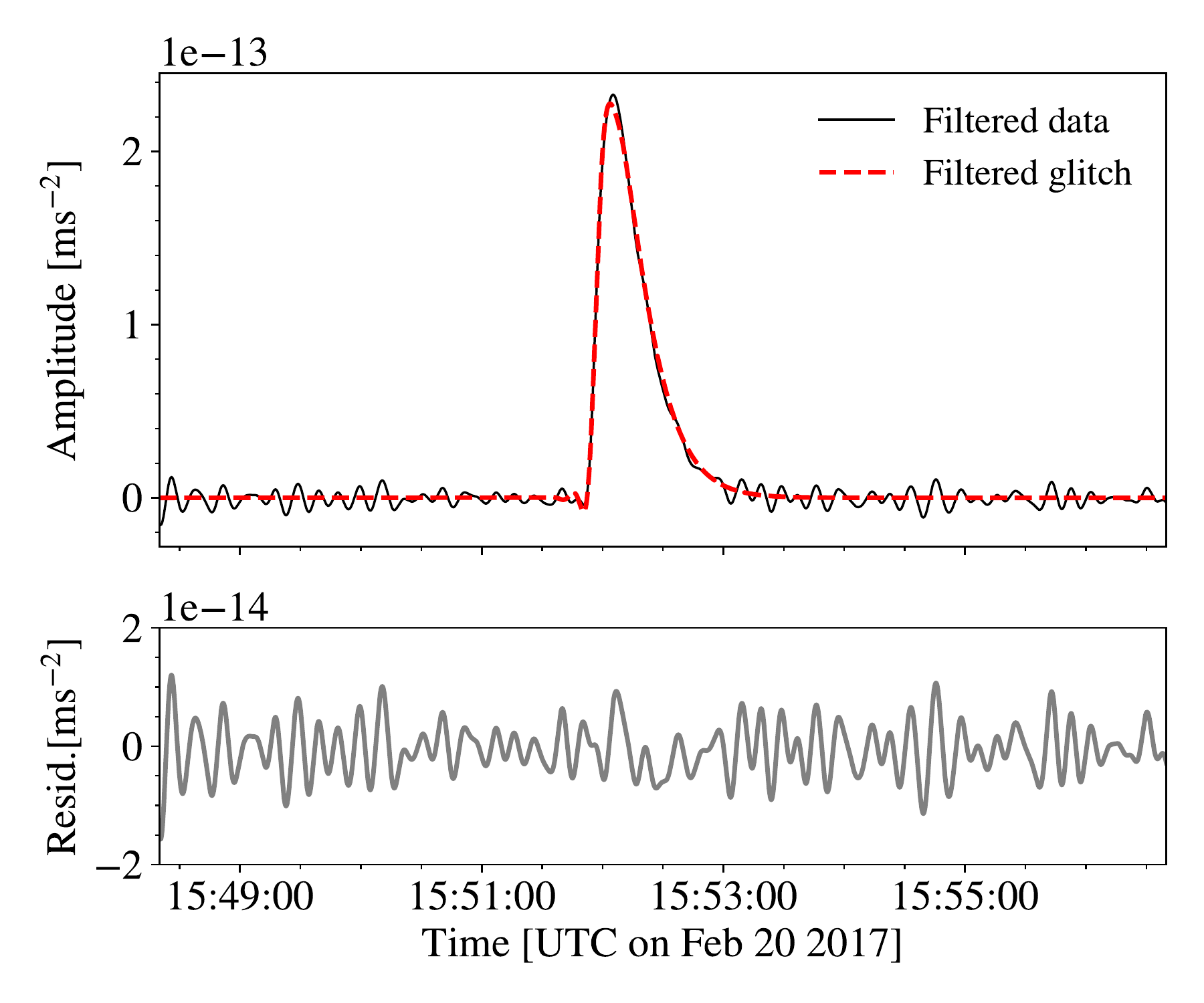} \\
		(a) FRED example \\
		\includegraphics[width=1.0\columnwidth,  trim={0.4cm, 0.3cm, 0.1cm, 0.3cm}, clip]{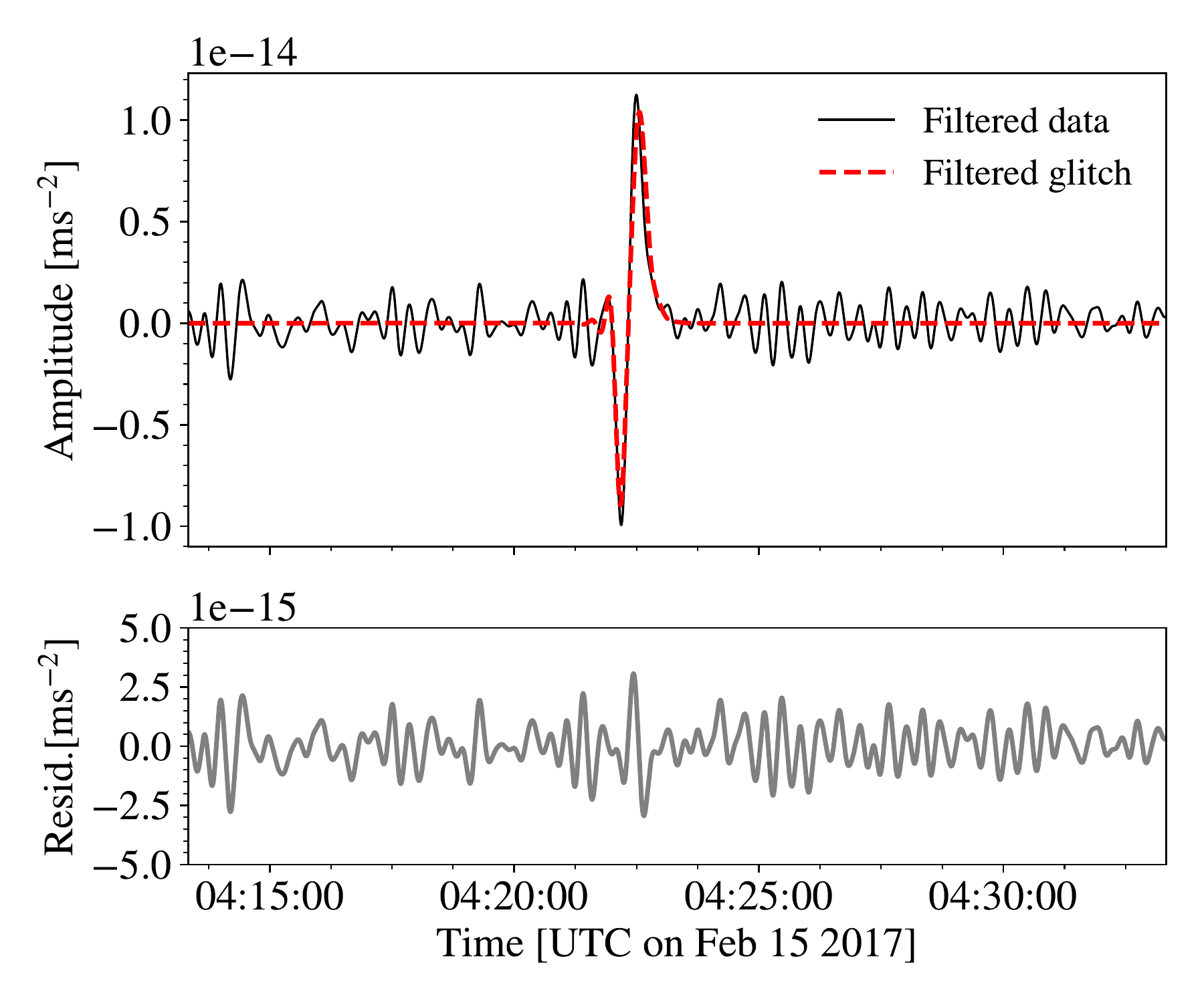} \\
		(b) Sine-Gaussian example
	\end{tabular}
	\caption{Excerpt of the noise run covering the period between February 13\textsuperscript{th} 2017 and March 2\textsuperscript{nd} 2017, zoomed on the first detection visible in Fig.~\ref{fig:snr_time_series_66}. The black solid line is the detrended $\Delta g$ measurement filtered using a Butterworth filter with cutting frequency 0.1 Hz~; the red dashed line is the fitted shapelet signal, and the grey curve is the fit residuals. }%
	\label{fig:glitch_fit_run_66_zoom}%
\end{figure}

We plot in Fig.~\ref{fig:glitch_fit_run_66_zoom} two examples of spotted events. The first one is a FRED event which corresponds to the SNR peak exhibited in the upper panel of Fig.~\ref{fig:snr_time_series_66}. We also plot a sine-Gaussian event occurring earlier in the run, corresponding to the peak shown in the lower panel of Fig.~\ref{fig:snr_time_series_66}.

\begin{figure}[!ht]
\centering
\includegraphics[width=1.0\columnwidth,  trim={0.5cm, 0.3cm, 0.1cm, 0.4cm}, clip]{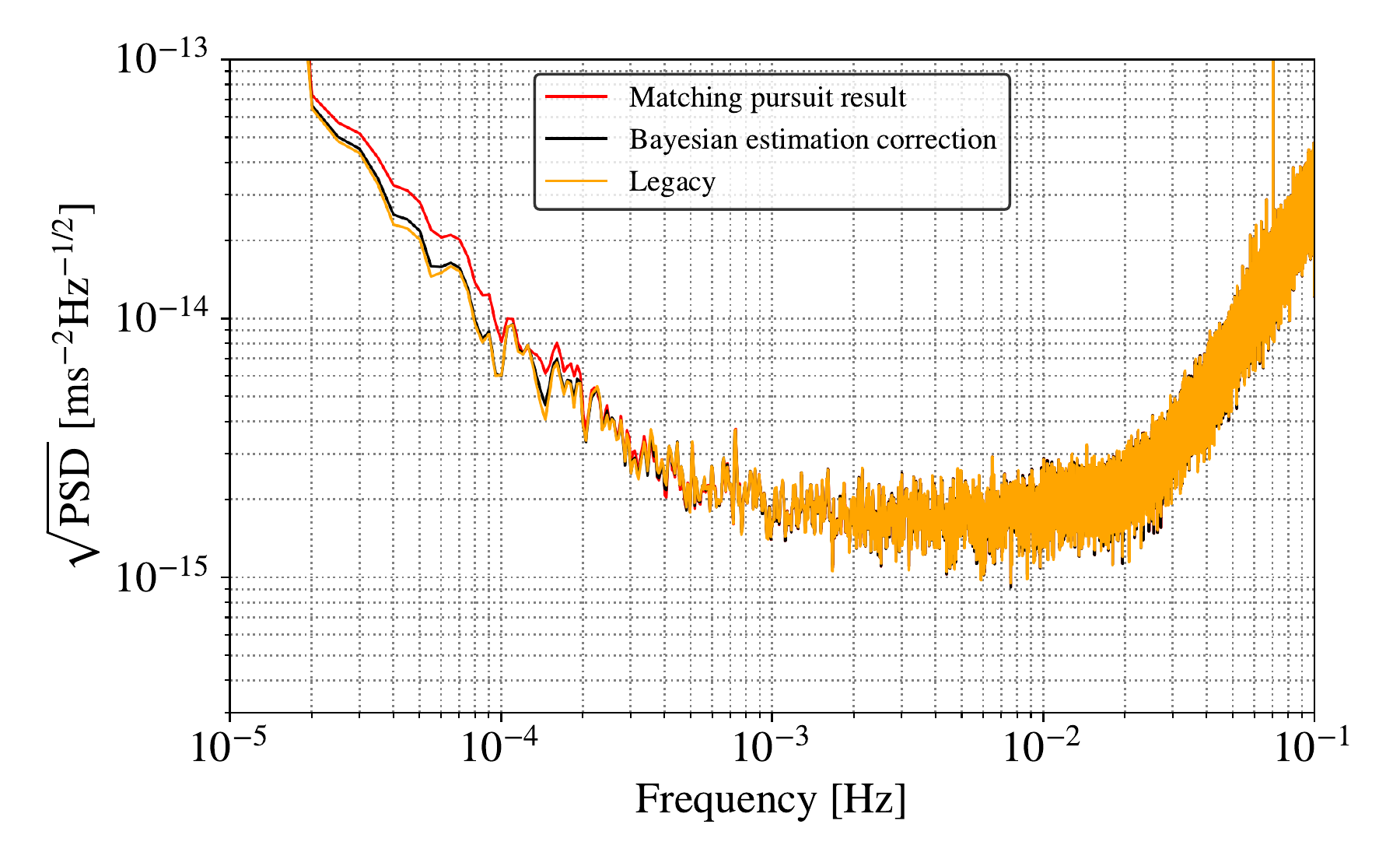}
\caption{Comparison of the shapelet fit residuals with the results of Ref.~\cite{lpf_prl2}. The red curve corresponds to the output of the matching filter algorithm, while the black curve is obtained by refining the glitch parameter estimates with an exploration of the posterior. The orange curve show the residuals of the fit that was performed in Ref.~\cite{lpf_prl2}.}%
\label{fig:residuals}%
\end{figure}

In addition, to cross-check the arrival times with the events spotted in the best performance noise run of February 2017, in Fig.~\ref{fig:residuals} we compare the shapelet fit residuals with the de-glitched data that are shown in Ref.~\cite{lpf_prl2}.
We plot the matching pursuit shapelet residuals (red) against Ref.~\cite{lpf_prl2}'s data (orange). We observe a slight excess of power at low frequency, below 0.2 mHz. This difference is due to the grid of damping parameter $\beta$ used in the matching pursuit search which is not fine enough to provide the best fit of the glitch signal, although it is sufficient to yield a reliable detection. This is particularly true for long-lived glitches where the maximum likelihood value lies between two grid elements. To improve the fit, we can refine the parameters estimation by sampling their posterior distribution using a Markov chain Monte-Carlo (MCMC) technique, as described in Section~\ref{sec:pe}. Starting with the matching pursuit outputs, we can quickly converge toward a more optimal value. By performing such a refinement on the longest events, we obtain the black residuals in Fig.~\ref{fig:residuals} which are close to Ref.~\cite{lpf_prl2}'s data within error bars. Note that this comparison only provides a consistency check between two deglitching approaches, since we do not have access to pure noise residuals.

\subsection{\label{sec:pe}Refined parameter estimation}

In this section, we check the results of the matched filtering process on selected events by performing a refined parameter estimation.
To this aim, we use a Bayesian approach by sampling the posterior distribution of the glitch parameters:
\begin{equation}
	p(\theta | x) = \frac{p(x | \theta) p(\theta) }{p(x)},
\end{equation}
where $p(x | \theta)$ is the likelihood function, $p(x)$ is the evidence and $p(\theta)$ is the prior distribution of the model parameters.
We estimate the posterior distribution for glitch parameters $\theta_s = \left( \alpha, \beta, \tau \right)$.

The prior distribution for all glitch parameters are chosen to be uniform around the values output by the matched filtering.
To probe the posterior, we use \textsc{ptemcee}~\cite{vousden_dynamic_2016, foreman-mackey_emcee_2013}, a stochastic sampling method based on MCMC with parallel tempering and linear transformations among chain states as proposals. We choose the longest glitch present in the noise run 66 as an example (see table~\ref{tab:segments}), as this event has a significant impact on the low-frequency part of the residual spectrum shown in Fig.~\ref{fig:residuals}. 

\begin{figure}[!ht]
	\includegraphics[width=\columnwidth, trim={0.3cm, 0.5cm, 0.3cm, 0.3cm}, clip]{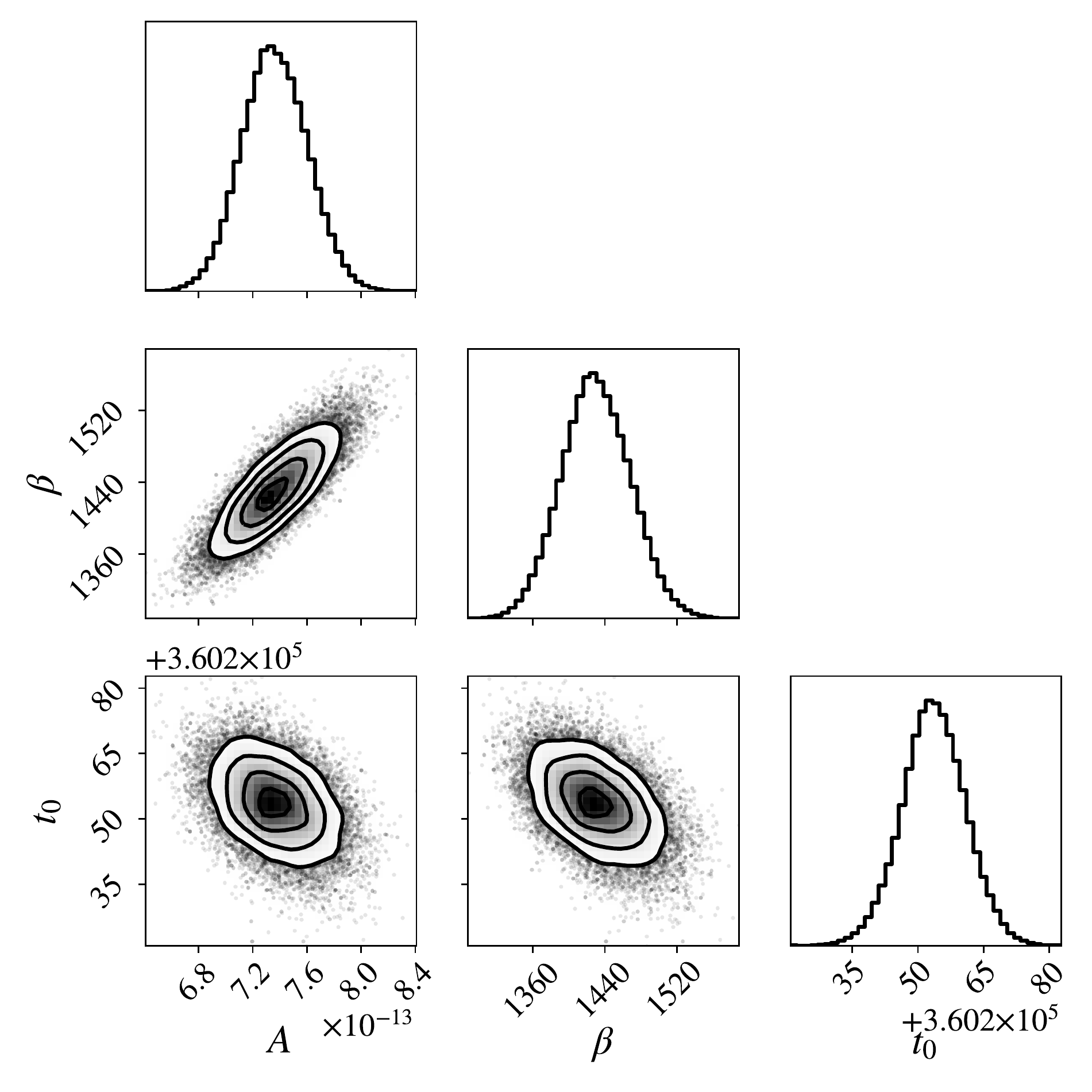}
	\caption{\label{fig:posterior}Posterior distribution of the amplitude ($\rm ms^{-2}$), damping time (s) and arrival time (s) for the longest glitch event in run 66. Priors are taken uniform around the matching pursuit output values. This plot was produced using the \textsc{corner} \textsc{python} package~\cite{corner}.}
\end{figure}
	
After sampling, the resulting posteriors for this investigation are shown in Fig.~\ref{fig:posterior}. Their marginal distribution follows closely the Gaussian probability density, and becomes stationary around the input values provided by the matching pursuit algorithms, which were {$\theta_s = \left(\SI{5.6e-13}{ms^{-2}}, \ \SI{1.096e3}{s}, \ \SI{3.6e5}{s} \right)$}.

Not surprisingly, the arrival time is the parameter best spotted by the matching pursuit detection, because the grid is as fine as the data sampling rate. The damping time grid being coarser, the Bayesian estimation allows one to better fit $\beta$ to an intermediate value. This has also an impact on the amplitude estimate $\alpha$, since it is correlated with $\beta$. The related improvement is visible on the residual plot shown in Fig.~\ref{fig:residuals}, where the noise level is lower at frequencies below 0.2 mHz.

This result advocates for including a glitch model as part of LISA global fit, i.e. the full characterization of all detectable sources in the data~\cite{littenbergglobal}. A glitch detection and subtraction as a pre-processing phase might not be enough to mitigate their impact on the noise PSD estimate and on the bias of gravitational waves (GW) source parameters.

\section{\label{sec:statistics}Analysis of glitch parameter statistics}


\subsection{Description of the observed population}

In this section, we analyse the distribution of glitch parameters among all detected events. Since some glitches are described by more than one basis component, we assume that flagged arrival times occurring within less than 5 seconds belong to the same event. We represent such multi-component glitches by an equivalent triad ($\alpha_{\mathrm{eq}}$, $\beta_{\mathrm{eq}}$,  $\tau_{\mathrm{eq}}$) corresponding to the component of maximum SNR among the sum of components describing the event. Note that due to the normalization of the shapelet function in Eq.~\eqref{eq:shapelet_model}, the effective amplitude of a glitch is proportional to $\alpha / \beta$. Therefore the parameter $\alpha_{\mathrm{eq}}$ corresponds to an equivalent transferred impulse $\alpha_{\mathrm{eq}} = \Delta v_{\mathrm{eq}}$ expressed in units of velocity (\si{\meter\per\second}). This quantity is important as it is directly related to the event SNR.

\begin{figure}[!ht]
	\centering
	\includegraphics[width=1.0\columnwidth,  trim={0.5cm, 0.3cm, 0.3cm, 0.3cm}, clip]{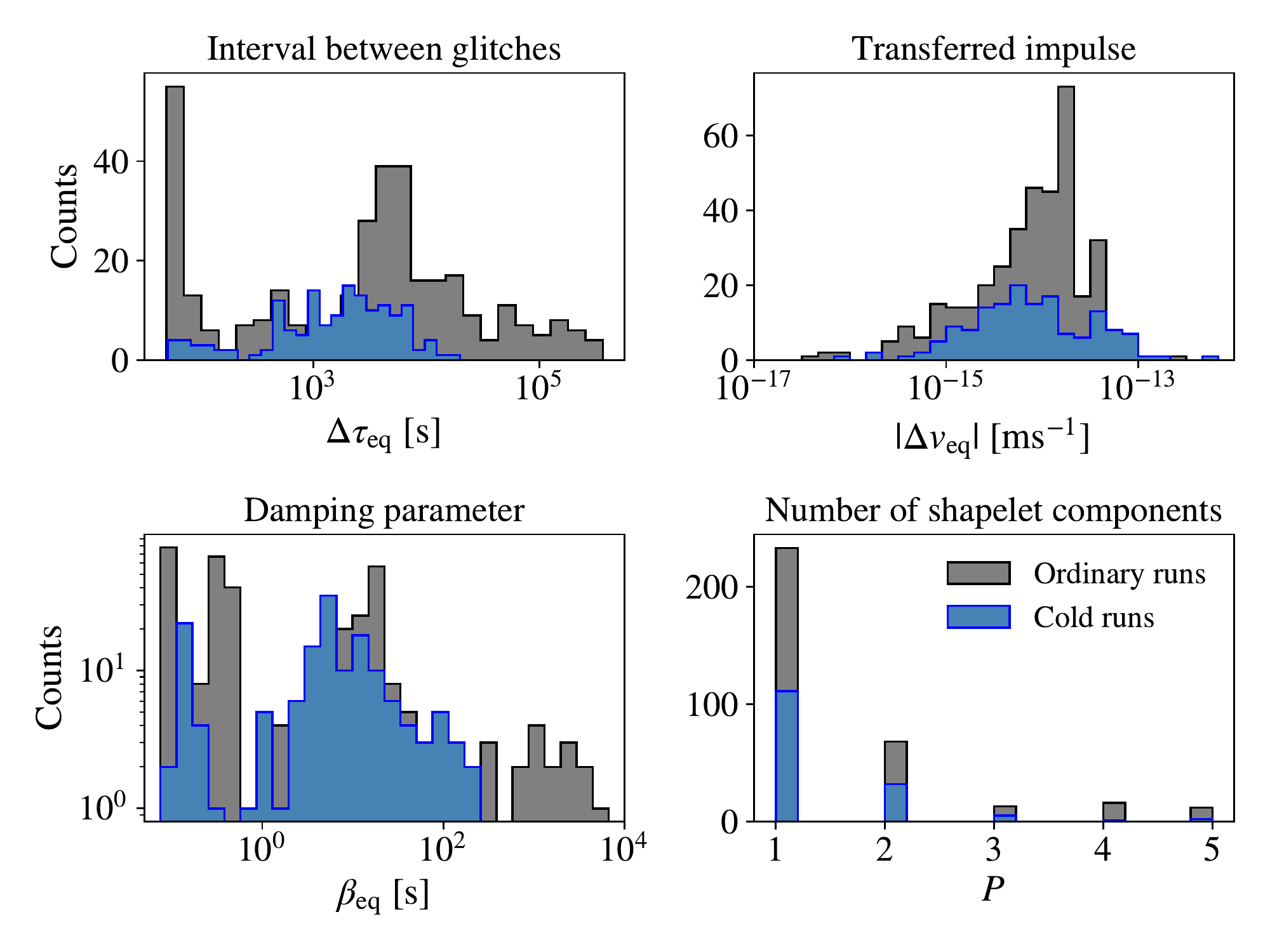}
	\caption{Histograms representing the distribution of equivalent glitch parameters from all detections, with logarithmic parameter bins. Top left panel: difference in arrival time between sequential glitches~; top right panel: equivalent transferred impulse; bottom left panel: equivalent damping time; bottom right panel: number of shapelet components per detected event. We distinguish cold (blue) and ordinary (gray) runs as they are affected by different rates of event occurrence.}%
	\label{fig:glitch_param_statistics}%
\end{figure}

We gather in Fig.~\ref{fig:glitch_param_statistics} the histograms of the parameters from all detected events. We separate the cold runs statistics (which represent 20\% of the number of events) from the ordinary runs statistics. The upper left panel shows the histogram of time intervals between glitches, which follows an exponential distribution. We estimate a glitch rate of $\lambda \approx 5\cdot 10^{-5}$ s$^{-1}$ (about 4 events per day) for ordinary runs, and a rate about ten times larger for cold runs. 

	The histograms in the upper right panel exhibit transferred impulses mostly ranging between $10^{-15}$ and $10^{-13}$~\si{\meter\per\second}. Note that the lower bound is likely to be due to the observation bias, i.e. the impossibility to detect events below the noise level. For example, for the estimated PSD in noise run 66, for $\beta \approx 1$ minute and a SNR detection threshold of 5, events with amplitudes below \SI{2E-16}{\meter\per\second} are not detected. 
	
	The lower left panel represents the damping times distributions, which show that most events last less than one minute, with rarer events lasting a few hours. The peaks visible at the leftmost edge of the damping time histograms are due to events whose timescales are either close or below the sampling time, resulting in Dirac-like signals.
	
	We also plot the statistics of the number of shapelet components needed to describe single events. In most cases, glitches have FRED shapes and one component is enough to describe them. However, some glitches require two or more components for their fit, like the sine-Gaussian signal shown in the lower panel of Fig.~\ref{fig:glitch_fit_run_66_zoom}.

\begin{figure}[!ht]
	\includegraphics[width=\columnwidth, trim={0.3cm, 0.5cm, 0.3cm, 0.5cm}, clip]{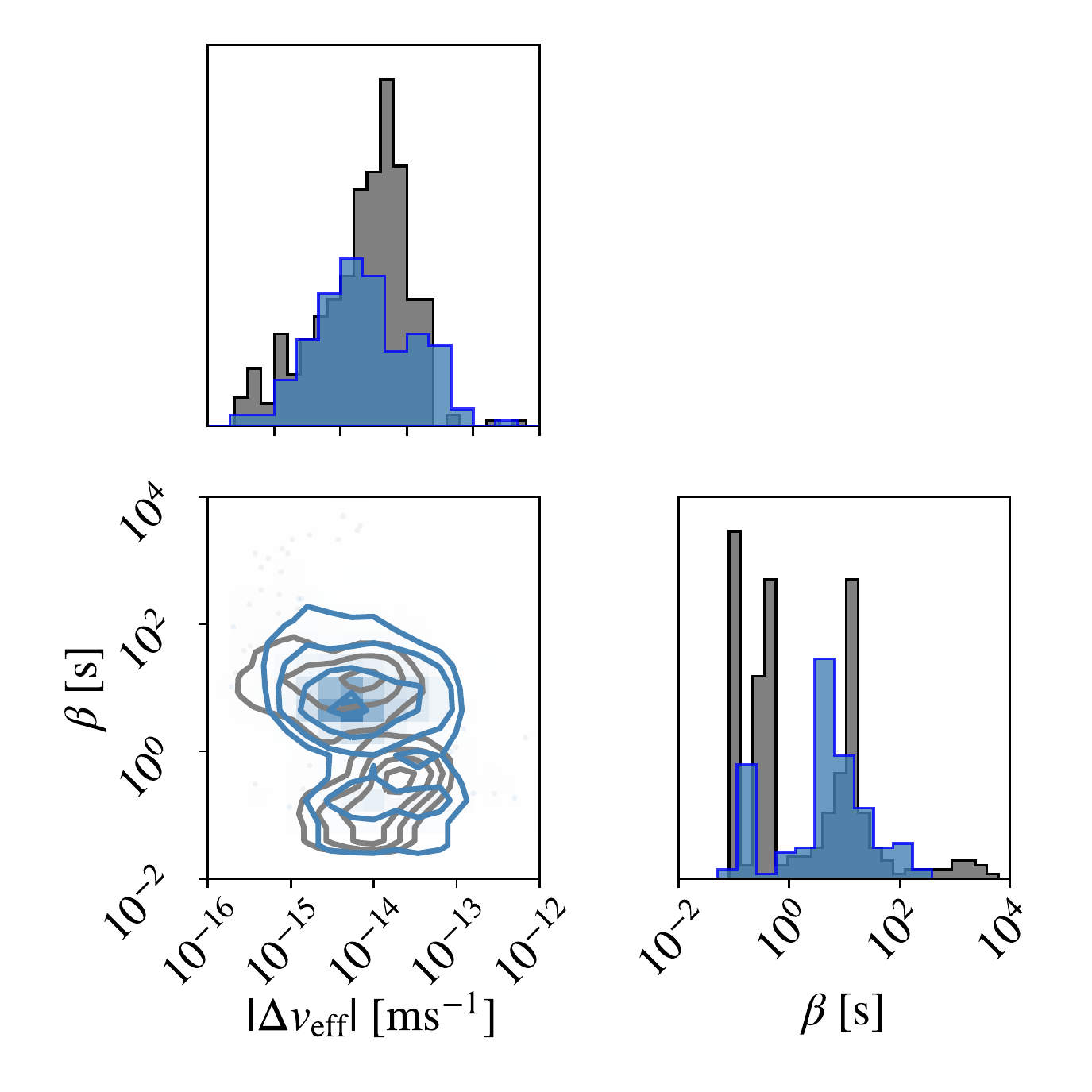}
	\caption{Smoothed contour plots for the distribution of the equivalent damping time $\beta_{\mathrm{eq}}$ (\si{\second}) and the transferred impulse $|\Delta v_{\mathrm{eq}}|$ (\si{\meter\per\second}, absolute value) in log scale. Blue histograms correspond to the samples from the {\it cold run} and gray histograms correspond to the samples from the {\it ordinary run}.\label{parameter_distribution}}
\end{figure}

To examine correlations among the glitch waveform parameters, we plot in Fig.~\ref{parameter_distribution} the joint distribution of $\beta_{\mathrm{eq}}$ and $\Delta v_{\mathrm{eq}}$ in the form of a corner plot. We can crudely distinguish two populations of events, especially when considering the ordinary runs (gray). The first category gathers short-duration events with damping times below one second (lower contours in the figure), while the second category includes longer events, with a peak around one minute (upper contours). The impulses of short events are on average slightly larger than the long ones. 

\subsection{Sampling from the glitch parameter distribution}

For future instrument and data analysis characterization efforts, it will be important to be able to generate as many glitch events as needed for LISA simulations. To this aim, we developed a method to sample the 2-dimensional distribution of the glitch amplitudes and damping times, to enable the corresponding LISA waveform generation. 

Drawing new samples representative of the LPF glitch population requires sampling the complicated two-dimensional distribution of the glitch parameters $(\alpha, \beta)$ presented in Fig.~\ref{parameter_distribution}, and sampling the times of arrival $\tau$. The approach that we adopt to draw samples from the desired distribution follows the idea that one can define an invertible map from the simple normal distribution $\mathcal{N}(\mathbf{0},\mathbf{I})$ to the desired distribution. This can be formalized using the following change of variable equation:
\begin{equation}
	p(\theta) = \mathcal{N}(z^{-1}(\theta)) \left| \text{det} \frac{\partial z^{-1}(\theta)}{\partial \theta} \right| ,
\end{equation}
where the parameters of the glitch are defined as ${\theta = (\alpha, \beta)}$, $p(\theta)$ is the distribution in question and $z(.)$ is the map that we want to estimate. The map can be estimated using machine learning techniques called {\it normalising flows}. We used a particular implementation called neural spline flows~\cite{durkan2019neural}.

We sample the glitches arrival times using the exponential distribution of the intervals between ordinary run events that we fitted from the upper left histogram in Fig.~\ref{fig:glitch_param_statistics}.

\section{\label{sec:glitch_generator}Extrapolating the glitch distribution from LPF to LISA}

Our analysis of LPF glitches can serve to inform a tentative assessment of the impact of spurious instrumental transients on LISA data processing. Another objective is to develop the tools needed to test data analysis techniques which will be robust to their presence. To do this assessment, we need a process to translate LPF measurement into equivalent LISA simulations. 

In LPF, the presence of glitch signals could not be detected in any measurement other than the most sensitive channel that was the differential interferometer measuring the position of one TM relative to the other. If we consider the same scenario in LISA, we have to assume that glitches may only be visible in the most sensitive channels as well, which are time-delay interferometry (TDI) observables~\cite{tinto_cancellation_1999, tinto_time-delay_2014}. TDI variables allow for the cancellation of laser frequency noise and are obtained from a post-process performed on ground. Besides, data analyses are generally performed in units of relative laser frequency deviations, also referred to as fractional frequency. Converting LPF data into equivalent relative laser frequency deviations requires integration with respect to time of the analytical shapelet model. Then, it needs to be projected to LISA arms and transformed through the TDI algorithm.
	
This approach was implemented in \textsc{LISA Glitch}~\cite{lisaglitch}, a Python module easily interfaced with the rest of the simulation tool chain. \textsc{LISA Glitch} is able to generate glitch signals of various shapes, including the shapelet model introduced in Sec.~\ref{sec:shapelets}. These glitch signals are associated with one injection point and written to a glitch file. This glitch file is then read by the instrument simulator and glitch signals are injected into their respective injection points during the simulation.

\subsection{Derivation of relative frequency deviations induced by glitches}

To adjust the glitch shapes to the commonly used units we first integrate the acceleration perturbation $\Delta a$ into a fractional frequency signal ${d\nu}/{\nu_0}$ as
\begin{equation}
	\label{eq:fractional_frequency_signal}
	\frac{d\nu}{\nu_0} = \int_{0}^{t} \frac{\Delta a(t')}{c} dt'.
\end{equation}
Plugging the shapelet model into the above equation yields (see Appendix~\ref{apx:lisa_response})
	\begin{equation}
	\label{eq:fractional_frequency_signal_shapelet}
	\frac{d\nu}{\nu_0}(t) =   \frac{2 \alpha}{c\beta} \left( 1  - e^{-\frac{t-\tau}{\beta}} \left( \frac{t-\tau}{\beta} + 1 \right) \right)h_{+}(t - \tau).
\end{equation}

\begin{figure}[!ht]
\centering
\begin{tabular}{c c}
\includegraphics[width=0.49\columnwidth]{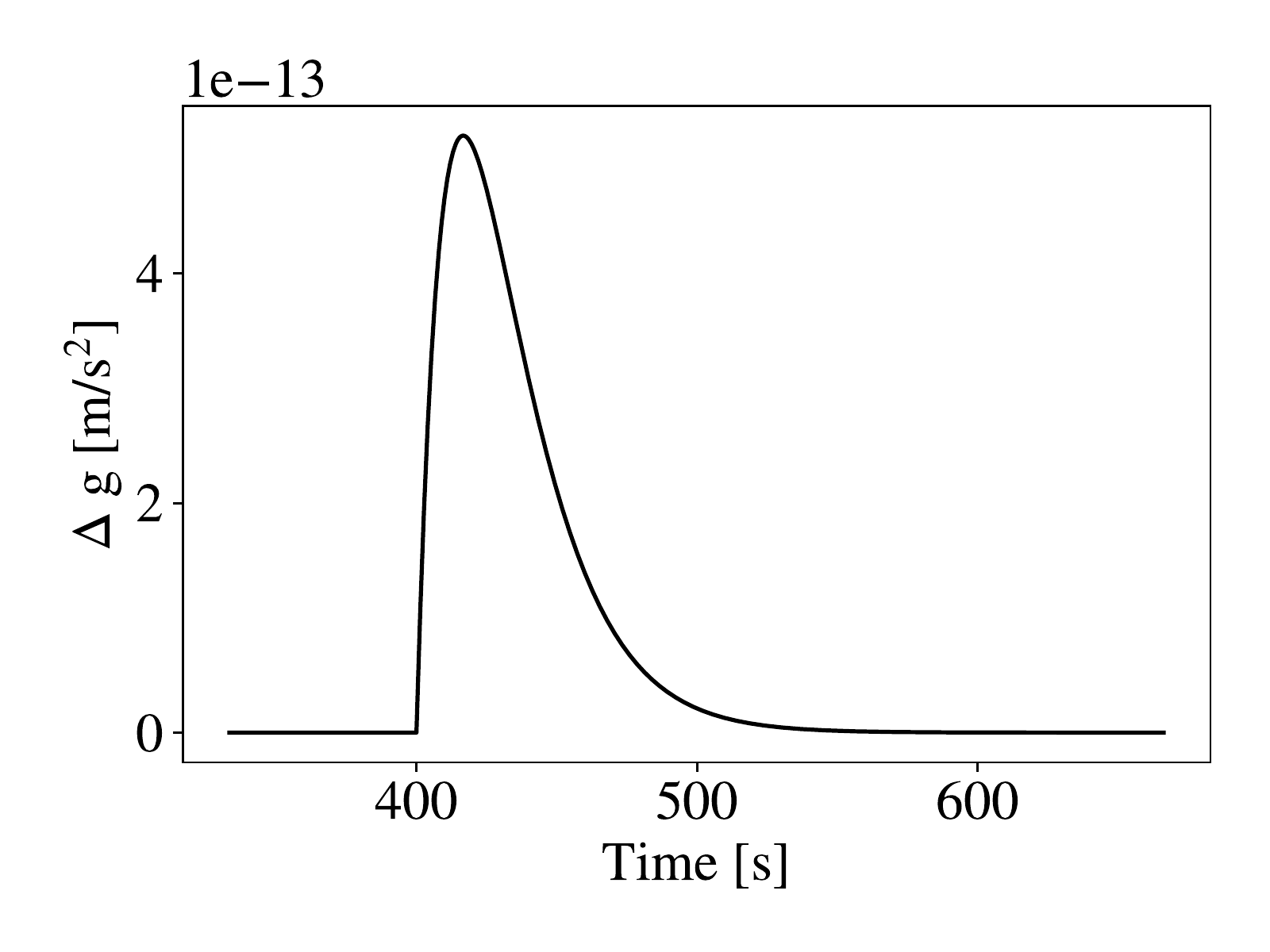} &
\includegraphics[width=0.49\columnwidth]{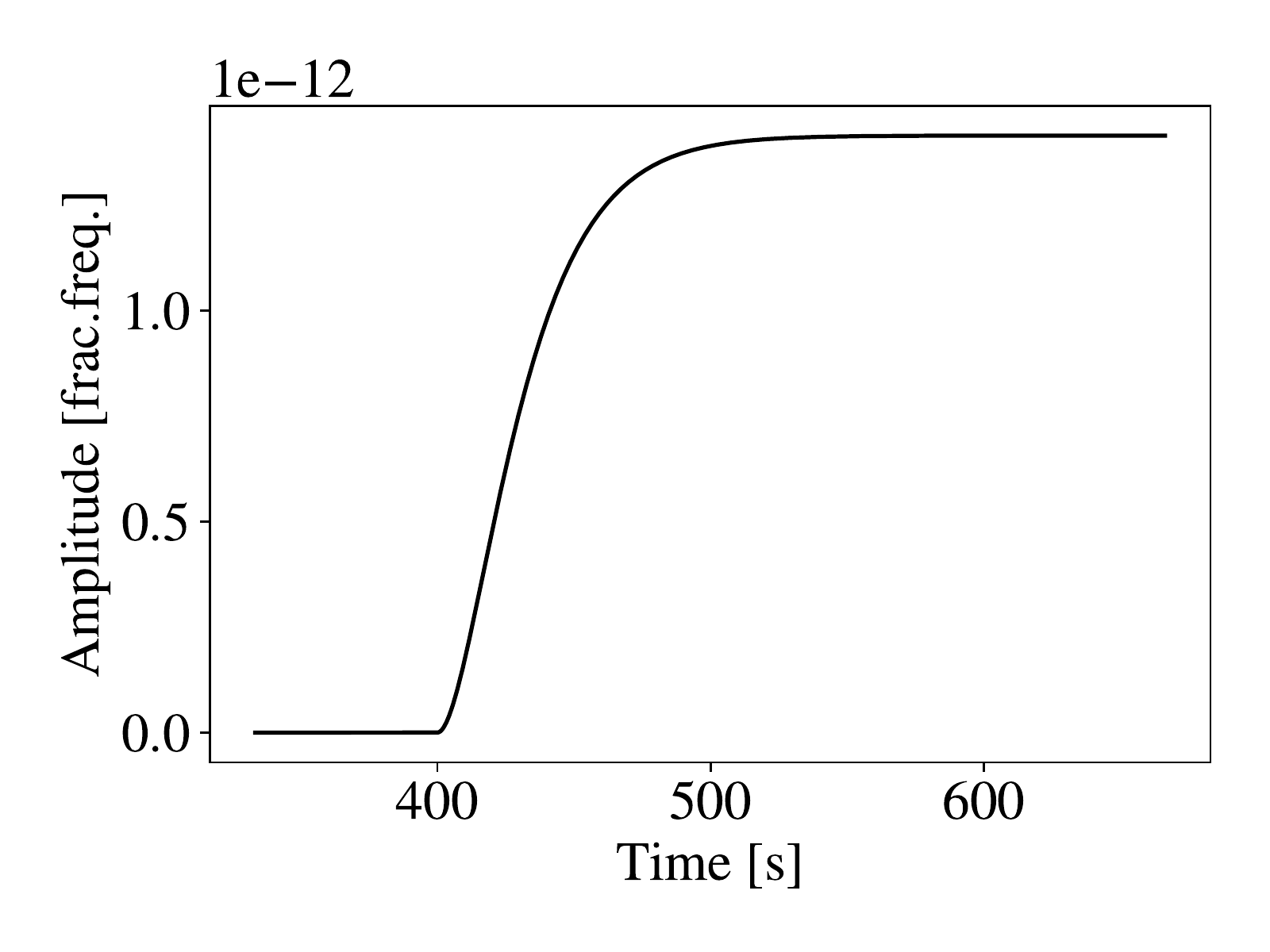} \\
(a) Differential acceleration \label{fig:glitch_acc}&
(b) Fractional frequency \label{fig:glitch_ff}
\end{tabular}
\caption{Generated glitch with the shapelet model for the parameter values $\beta = \SI{50}{\second}$ and $\alpha = \SI{5E-12}{\meter\per\squared\second}$ (panel (a)); and the representation of the same glitch in fractional frequency (panel (b)). }%
\label{fig:glitch_models}%
\end{figure}

Graphs of the original glitch in differential acceleration and the integrated glitch in fractional frequency are shown in Fig.~\ref{fig:glitch_models}.

\subsection{Time-delay interferometry response to glitches}
    
A thorough investigation of glitch propagation would require introducing them at different simulated injection points, propagating the signals through the LISA dynamics and observation of the output in different interferometers. We postpone such an in-depth study for the future, as the development of LISA dynamics simulations is still in progress at the time of the writing of this paper. In the absence of dynamics, we inject glitches directly on the {\it test-mass} interferometer or on the inter-spacecraft ({\it science}) interferometer. We look at how the resulting output in the TDI channels will differ based on the place of injection.
In reality the appearance of the glitches in one of the interferometers will depend on their frequency content and on the Drag Free and Attitude Controllers (DFACS)~\cite{dfacs} frequency response.

 \begin{figure}[!ht]
\includegraphics[width=0.48\columnwidth]{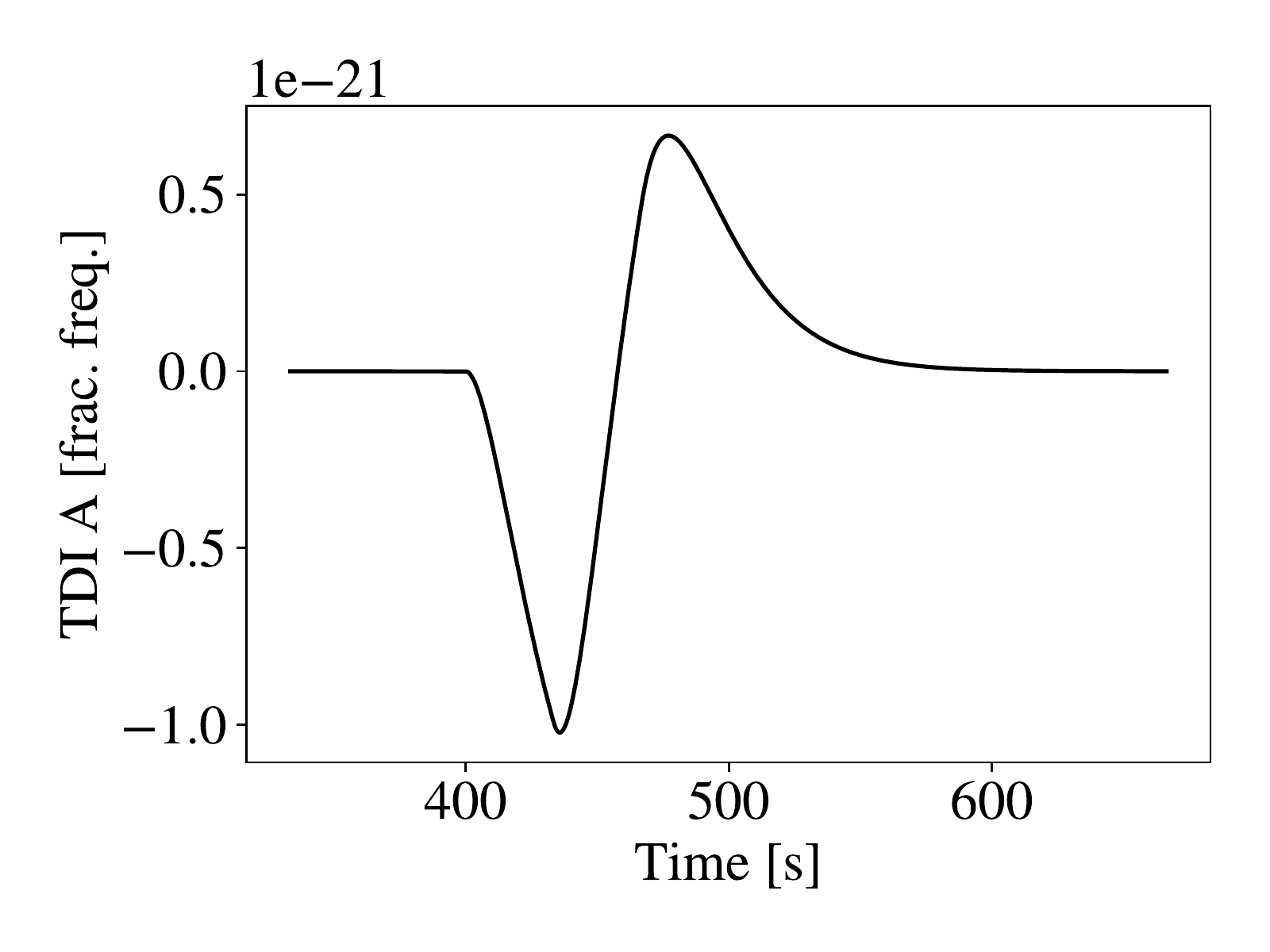}   \includegraphics[width=0.48\columnwidth]{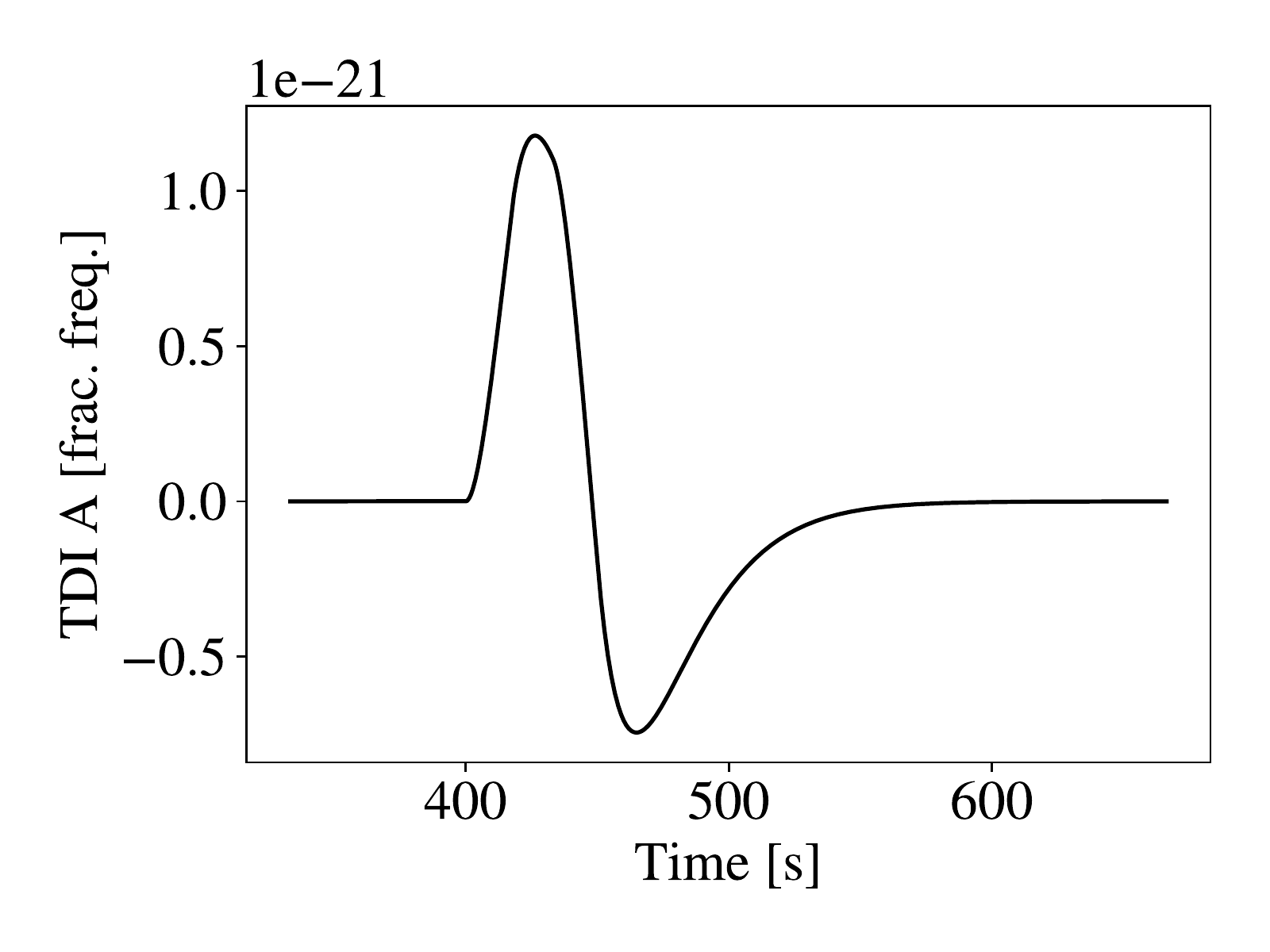}  \\
\includegraphics[width=0.48\columnwidth]{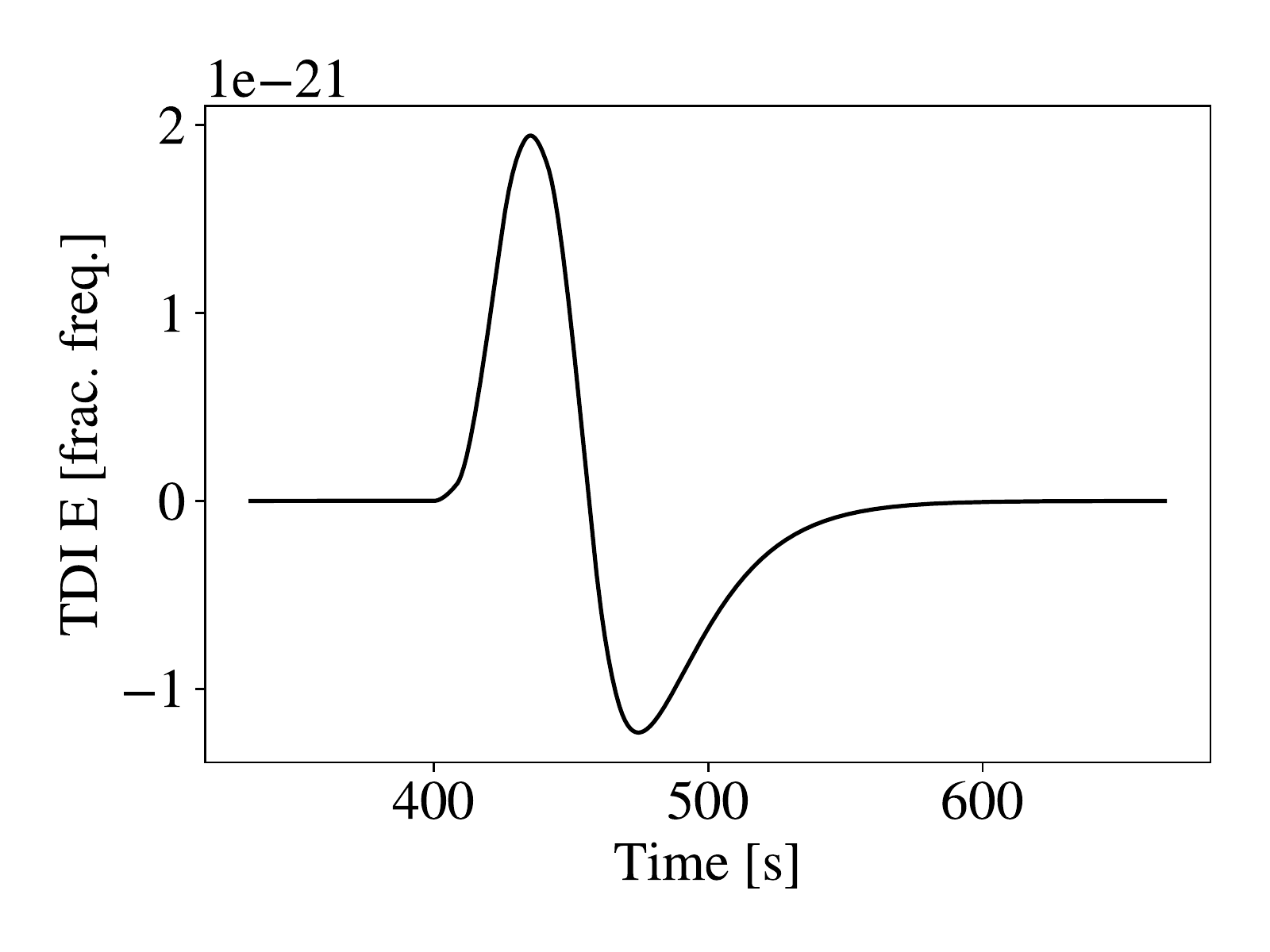}  \includegraphics[width=0.48\columnwidth]{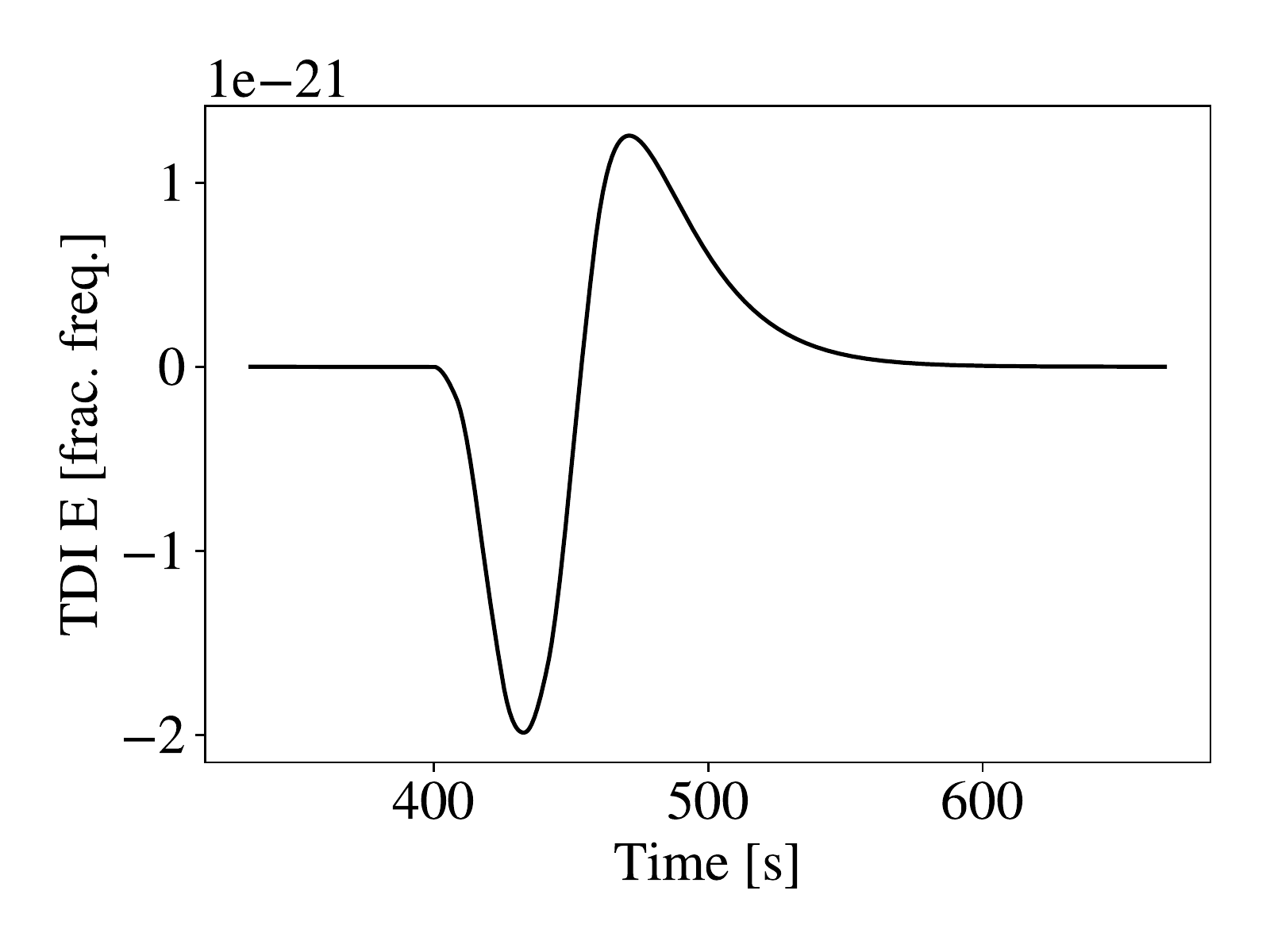}  \\
 \includegraphics[width=0.48\columnwidth]{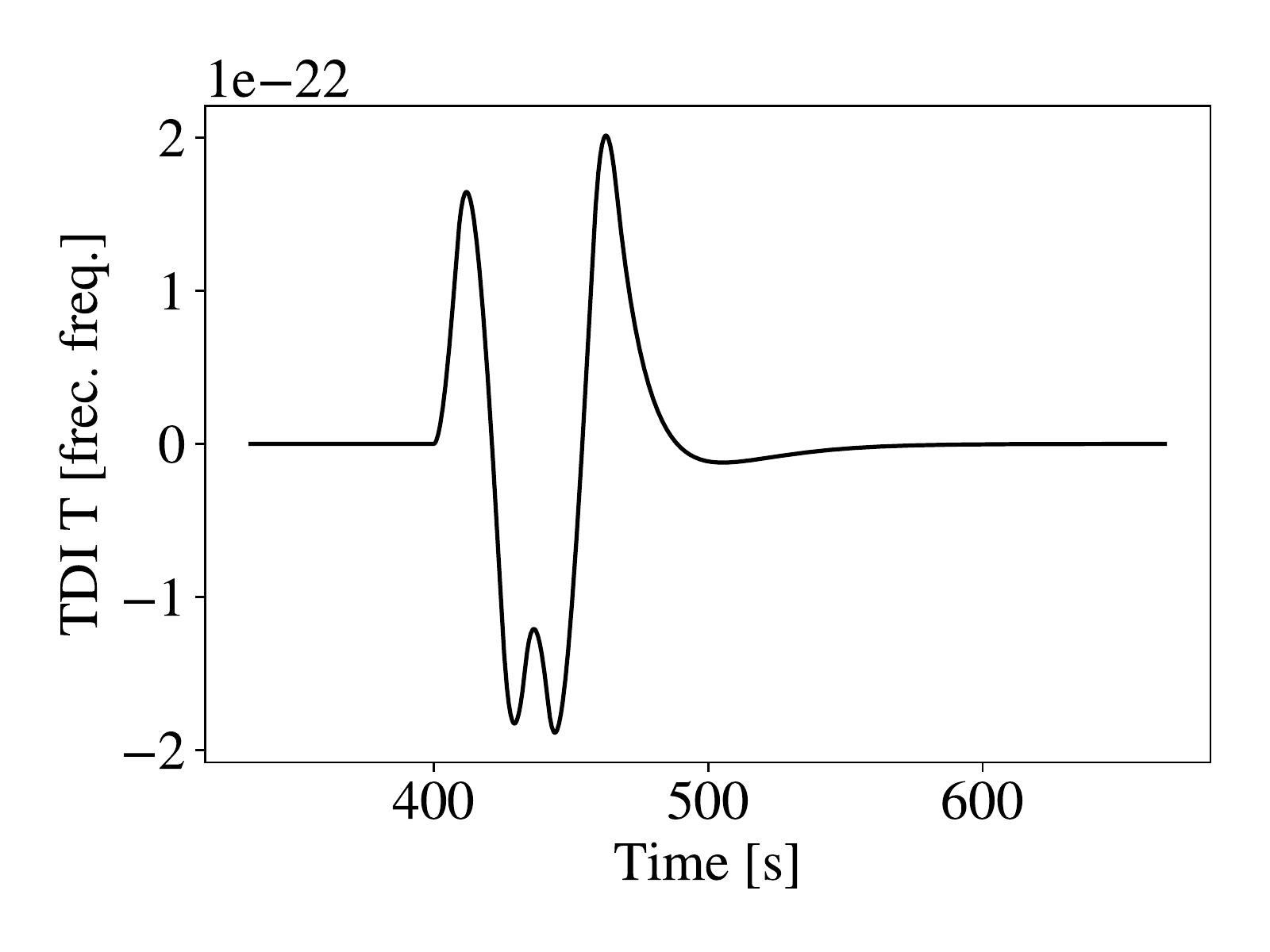}   \includegraphics[width=0.48\columnwidth]{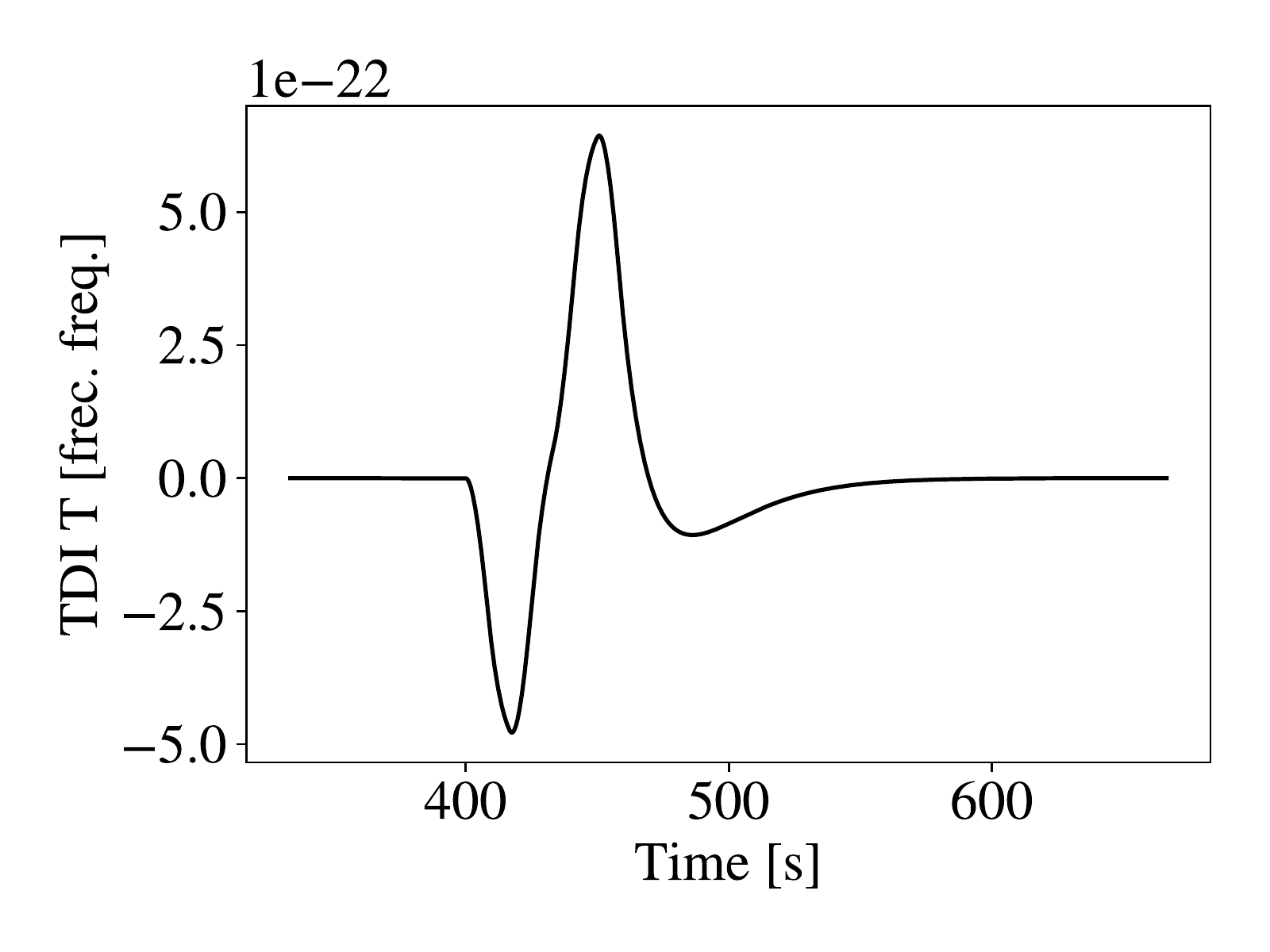} 
\caption{\label{fig:tdi_propagation}
Glitch response in TDI combinations $A$ (first row), $E$ (second row), and $T$ (third row) for an injection at the {\it test-mass} (left column) and {\it science} (right column) interferometers.}
\end{figure}

We consider the triple quasi-orthogonal combinations $A$, $E$ and $T$ which are typically used in the literature~\cite{prince_lisa_2002}. The advantage of these combinations is that they exhibit quasi-orthogonal noise and $T$ is a ``null" combination which suppresses most of the GW signal at low frequency. To propagate the glitches through LISA data streams, we use the {\textsc pyTDI} package~\cite{pytdi} which computes TDI combinations from raw interferometer measurements. We show an example in Fig.~\ref{fig:tdi_propagation}. We see that the smooth step function shape at the interferometer level in Fig.~\ref{fig:glitch_models} is turned into sharper features in the TDI combinations, which act as a differentiation on a duration of a few inter-spacecraft delays (about 8 seconds). 
We inject the glitch signal into one single {\it science} interferometer out of six. Note that in reality, the reaction of the spacecraft drag-free system control loop may result in occurrences of the glitch signal in more than one interferometer. More accurate and realistic propagation of glitches in LISA is the subject of an on-going study.

\subsection{Glitch SNR in LISA}

Using the simulation described in the previous section, we can compute the glitch TDI response, and hence the SNR in simulated LISA data. Although this computation relies on the assumptions we made regarding the injection point, it can provide us with a good order of magnitude estimate. In the following, we choose to treat glitches as a perturbation on one of the TMs.

We have to assume a noise model for the second-generation pseudo-orthogonal TDI channels, reflecting the main pathlength noises in the measurement. We use the model provided by the \textsc{LDC} software~\cite{LDC}, based on the LISA science requirement document~\cite{lisa_simulation_working_group_lisa_2018}. 

\begin{figure}[!ht]
\includegraphics[width=0.98\columnwidth]{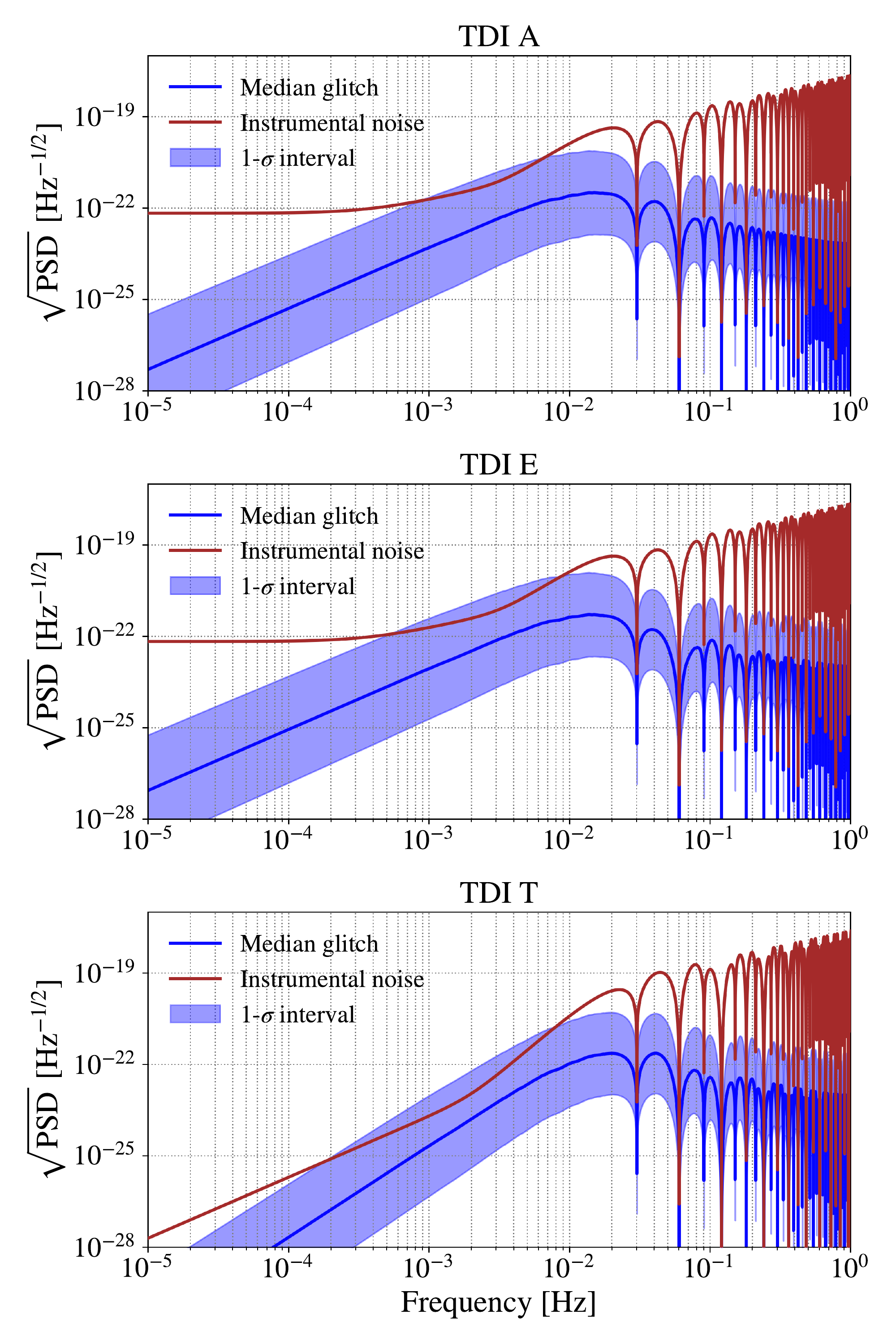} 
\caption{\label{fig:tdi_psds}Median (blue) and 1$\sigma$-confidence interval (light blue area) of the distribution of all detected glitches PSDs in the TDI channel $A$ (upper), $E$ (middle) and $T$ (lower), compared to the theoretical noise PSD (red) expressed in fractional frequency deviation.}
\end{figure}

First, we compare the glitch power in the TDI combinations $A$, $E$, and $T$ with the theoretical PSD of the stochastic instrumental noise. We choose an observation time of $\SI{E5}{\second}$, which encompasses the duration of the longest glitch, and compute the empirical PSDs of all detected events. We plot the median and the standard deviation of this PSD distribution in Fig.~\ref{fig:tdi_psds}, along with the noise PSD.

We see that the median of the glitch power lies below the mean noise level in the LISA frequency band. However, the spreading of the power across a large bandwidth implies large SNRs. Integrating the distribution over all frequencies yields the events SNRs in TDI channels:
\begin{eqnarray}
	\label{eq:lisa_snr}
	\mathrm{SNR}^2_{\mathrm{LISA}} = \sum_{i = A, E, T} \int_{f_{\mathrm{min}}}^{f_{\mathrm{max}}} 4 \frac{\lvert \tilde{s}_{i}(f) \rvert^2}{S_{i}(f)}df,
\end{eqnarray}
where $\tilde{s}_{i}(f)$ is the Fourier transform of the glitch response in TDI $i$, and $S_{i}(f)$ is the instrumental noise one-sided PSD in channel $i$. We restrict the integration to frequencies greater than $f_{\mathrm{min}} = 10^{-5}$ Hz and lower than $f_{\mathrm{max}} = 1$ Hz, which are the boundaries of the observatory sensitivity goal, as stated in the LISA Science Requirement Document~\cite{lisa_simulation_working_group_lisa_2018}.
	
 \begin{figure}[!ht]
	\includegraphics[width=0.98\columnwidth]{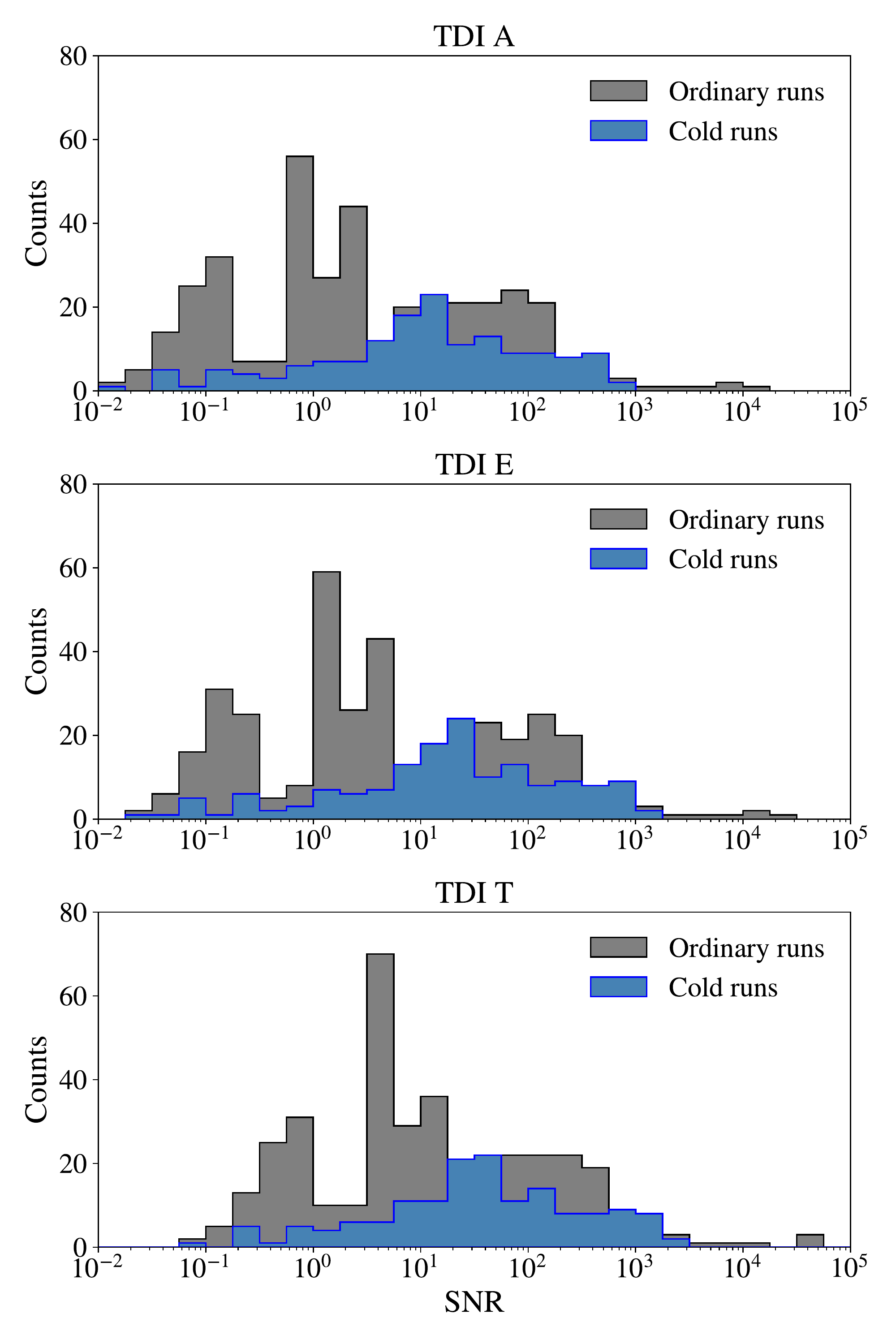}   \\
	\caption{\label{fig:lisa_snrs}Distribution of extrapolated SNRs from LPF events in the orthogonal TDI $A$ (upper panel), $E$ (middle), and $T$ (bottom) channels, for ordinary (gray) and cold (blue) runs, assuming a single perturbation in one of the TMs.}
\end{figure}

Fig.~\ref{fig:lisa_snrs} shows the distribution of TDI SNRs derived from LPF events detected in ordinary (gray) and cold (blue) runs. It exhibits values that range between $10^{-2}$ to $10^4$, with 50\% of events having SNRs larger than 10. The maximal values are large with respect to typical GW sources, which can reach a few hundreds for galactic binaries to a few thousands for supermassive black holes. However, glitch signals are very short in time so they can easily be isolated in time-frequency representations. This very preliminary analysis suggests the need for an adequate processing to mitigate the impact of glitches on science performance.

\section{Conclusion}
	
In this paper, we investigated the presence of transient noise artefacts in the LPF noise runs. We developed a method to detect these glitches based on the matched filtering of shapelets, a family of adapted mathematical functions. Each shapelet component is parametrized by an amplitude and a decay time, which can be interpreted as an equivalent transferred impulse amplitude and duration, respectively.

We characterize the glitch population by applying the detection scheme to 56 differential acceleration measurements, from which we obtain a distribution of amplitudes, damping times, and arrival times. Then, we develop a machine-learning method that allows one to draw an arbitrary number of events from this distribution. 

We explore possible consequences for LISA by making assumptions on glitch injection points, i.e. on the system components from which they may originate. Based on these assumptions, we sketch a preliminary process to compute equivalent perturbations in LISA TDI data. 
We find that half of the detected events have a significant SNR when extrapolated to LISA measurements. This result makes it all the more necessary to study in depth the impact of transient disturbances on LISA performance.

The tool that we develop constitutes a first building block to inform studies assessing the impact of glitches on LISA scientific performance, and develop adapted data analysis methods to mitigate it. The outputs of this study are currently used in the LISA Data Challenge dataset codenamed \textit{Spritz}~\cite{LDC}, where we simulate LISA data corrupted by LPF-like glitches. Investigating the effect of glitches on the detection and characterization of GW sources is paramount for the LISA mission development. While some studies already tackled the problem of distinguishing instrumental glitches from GW bursts~\cite{Robson2019}, more realistic simulation campaigns including LPF-like signals and dynamical instrumental response will be carried out in the near future.

\begin{acknowledgments}
We would like to gratefully thank the members of the LISA Consortium Simulation and the LISA Data Challenge working groups for interesting inputs and exchanges. We also thank the LISA Pathfinder Collaboration for generating the delta-g data products and making them available to the community. This work is carried out as a contribution to the {\it Artifacts Group} within the LISA Science Group Work Package 2 (WP2). We warmly thank Ira Thorpe, Joseph Martino and Antoine Petiteau for their valuable feedback. N. Korsakova acknowledges the support from CNES fellowship and the support by the LABEX Cluster of Excellence FIRST-TF (ANR-10-LABX-48-01), within the Program "Investissements d'Avenir” operated by the French National Research Agency (ANR). Q. Baghi acknowledges the support from an appointment to the NASA Postdoctoral Program at the Goddard Space Flight Center, administered by Universities Space Research Association under contract with NASA. He also thanks Nelson Christensen for hosting him in the ARTEMIS lab at C\^{o}te d'Azur Observatory to enable a collaboration on this work.
\end{acknowledgments}


\appendix

\section{List of analyzed noise-only measurements}
\label{apx:list_segments}

We insert in Table~\ref{tab:segments} the list of the analyzed data segments.\\
	
	\begin{table*}[htbp]
		\caption{Differential acceleration noise runs used in this work.}
		\label{tab:segments}
		\rowcolors{2}{}{lightgray}
		\renewcommand*{\arraystretch}{1.1}
		\begin{tabular}{ccc}
			\hline
			{LPF Run index} & {Start time} & {Stop time} \\
			\hline
			1  &  \texttt{2016-03-01 08:05:00 UTC} & \texttt{2016-03-03 08:00:00 UTC} \\
			2  &  \texttt{2016-03-03 09:00:00 UTC} & \texttt{2016-03-05 08:00:00 UTC} \\
			3  &  \texttt{2016-03-06 18:15:00 UTC} & \texttt{2016-03-11 07:59:46 UTC} \\
			4  &  \texttt{2016-03-13 16:30:00 UTC} & \texttt{2016-03-15 07:00:00 UTC} \\
			5  &  \texttt{2016-03-16 20:00:00 UTC} & \texttt{2016-03-19 07:59:47 UTC} \\
			6  &  \texttt{2016-03-21 20:00:00 UTC} & \texttt{2016-03-26 08:00:00 UTC}\\
			7  &  \texttt{2016-03-27 14:00:00 UTC} & \texttt{2016-03-28 08:00:00 UTC} \\
			7  &  \texttt{2016-03-29 80:00:00 UTC} & \texttt{2016-03-30 08:00:00 UTC} \\
			8  &  \texttt{2016-03-31 07:02:00 UTC} & \texttt{2016-04-02 08:00:00 UTC} \\
			9  &  \texttt{2016-04-04 15:00:00 UTC} & \texttt{2016-04-14 08:00:00 UTC}\\
			10 &  \texttt{2016-04-18 19:00:00 UTC} & \texttt{2016-04-18 22:00:00 UTC}\\
			12 &  \texttt{2016-04-26 08:00:00 UTC} & \texttt{2016-04-30 08:00:00 UTC}\\
			13 &  \texttt{2016-05-01 08:05:00 UTC} & \texttt{2016-05-02 23:55:00 UTC}\\
			14 &  \texttt{2016-05-03 04:20:00 UTC} & \texttt{2016-05-05 15:30:00 UTC}\\
			15 & \texttt{2016-05-13 00:50:00 UTC} & \texttt{2016-05-13 07:30:00 UTC} \\
			16 &  \texttt{2016-05-13 08:30:00 UTC} & \texttt{2016-05-14 08:00:00 UTC}\\
			17 &  \texttt{2016-05-16 00:00:00 UTC} & \texttt{2016-05-19 08:00:00 UTC}\\
			18 &  \texttt{2016-05-19 08:30:00 UTC} & \texttt{2016-05-21 11:00:00 UTC} \\
			21 &  \texttt{2016-05-26 17:00:00 UTC} & \texttt{2016-05-27 01:00:00 UTC} \\
			37 &  \texttt{2016-06-15 13:30:00 UTC} & \texttt{2016-06-18 08:00:00 UTC}\\
			39 &  \texttt{2016-06-19 11:00:00 UTC} & \texttt{2016-06-25 08:00:00 UTC}\\
			40 &  \texttt{2016-07-10 08:00:00 UTC} & \texttt{2016-07-11 09:55:00 UTC}\\
			42 &  \texttt{2016-07-17 12:00:00 UTC} & \texttt{2016-07-20 06:00:00 UTC}\\
			43 &  \texttt{2016-07-24 07:40:00 UTC} & \texttt{2016-07-30 00:00:00 UTC}\\
			44 &  \texttt{2016-07-31 10:10:00 UTC} & \texttt{2016-08-02 06:00:00 UTC}\\
			45 &  \texttt{2016-08-07 07:45:00 UTC} & \texttt{2016-08-08 04:20:00 UTC}\\
			47 &  \texttt{2016-08-16 13:15:00 UTC} & \texttt{2016-08-20 05:45:00 UTC} \\
			48 &  \texttt{2016-08-23 14:00:00 UTC} & \texttt{2016-08-27 20:00:00 UTC} \\
			49 &  \texttt{2016-09-05 11:35:00 UTC} & \texttt{2016-09-06 05:05:00 UTC} \\
			50 &  \texttt{2016-09-07 03:09:07 UTC} & \texttt{2016-09-07 04:50:00 UTC} \\
			51 &  \texttt{2016-09-11 21:15:00 UTC} & \texttt{2016-09-16 05:15:00 UTC} \\
			52 &  \texttt{2016-09-16 18:36:54 UTC} & \texttt{2016-09-17 08:00:40 UTC} \\
			53 &  \texttt{2016-09-19 01:30:00 UTC} & \texttt{2016-09-21 13:00:00 UTC}\\
			53 &  \texttt{2016-09-21 13:45:00 UTC} & \texttt{2016-09-22 06:00:00 UTC}\\
			54 &  \texttt{2016-09-28 13:35:00 UTC} & \texttt{2016-10-01 08:00:00 UTC}\\
			54 &  \texttt{2016-09-28 13:35:00 UTC} & \texttt{2016-10-01 08:00:00 UTC}\\
			55 &  \texttt{2016-10-04 00:00:00 UTC} & \texttt{2016-10-05 07:10:52 UTC} \\
			56 &  \texttt{2016-10-05 17:30:00 UTC} & \texttt{2016-10-07 00:49:00 UTC}\\
			56 &  \texttt{2016-10-07 02:15:00 UTC} & \texttt{2016-10-08 07:55:00 UTC}\\
			57 &  \texttt{2016-10-21 11:30:00 UTC} & \texttt{2016-10-21 16:39:00 UTC} \\
			58 &  \texttt{2016-11-07 14:20:00 UTC} & \texttt{2016-11-12 08:00:00 UTC}\\
			59 &  \texttt{2016-11-16 11:05:00 UTC} & \texttt{2016-11-26 08:00:00 UTC}\\
			60 &  \texttt{2016-12-02 23:35:00 UTC} & \texttt{2016-12-04 16:35:00 UTC} \\
			61 &  \texttt{2016-12-26 03:30:00 UTC} & \texttt{2017-01-13 19:58:00 UTC}\\
			63 &  \texttt{2017-01-27 11:02:00 UTC} & \texttt{2017-01-28 08:00:00 UTC}\\
			64 &  \texttt{2017-02-02 07:00:00 UTC} & \texttt{2017-02-02 20:20:00 UTC}\\
			66 &  \texttt{2017-02-13 08:00:00 UTC} & \texttt{2017-03-03 22:00:00 UTC}\\
			67 &  \texttt{2017-03-09 19:20:00 UTC} & \texttt{2017-03-14 09:40:00 UTC}\\
			68 &  \texttt{2017-03-14 09:00:00 UTC} & \texttt{2017-03-17 00:30:00 UTC}\\
			69 &  \texttt{2017-04-21 14:15:00 UTC} & \texttt{2017-04-24 07:45:00 UTC} \\
			71 &  \texttt{2017-05-03 23:30:00 UTC} & \texttt{2017-05-09 14:00:00 UTC}\\
			72 &  \texttt{2017-05-10 11:11:20 UTC} & \texttt{2017-05-12 12:02:07 UTC}\\
			73 &  \texttt{2017-05-12 12:02:17 UTC} & \texttt{2017-05-15 08:00:59 UTC}\\
			74 &  \texttt{2017-05-18 18:24:46 UTC} & \texttt{2017-05-23 02:00:00 UTC}\\
			75 &  \texttt{2017-05-28 13:41:00 UTC} & \texttt{2017-06-05 15:04:40 UTC}\\
			76 &  \texttt{2017-06-08 20:04:45 UTC} & \texttt{2017-06-17 02:56:00 UTC}\\
		\end{tabular}
	\end{table*}

\section{Derivation of LISA response to glitches}
\label{apx:lisa_response}
In this section, we derive LISA response to glitches in fractional frequency deviations, assuming they can be described by shapelets. For a shapelet of order $n = 1$, Eqs.~(\ref{eq:shapelet_model}) and Eq.~(\ref{eq:linear_combination}) give
\begin{eqnarray}
	\label{eq:shapelet_model_n=1}
	\Delta a(t) = 2 \alpha \frac{t-\tau}{\beta} e^{-\frac{t-\tau}{\beta}} h_{+}\left(\frac{t - \tau}{\beta}\right).
\end{eqnarray}
Plugging this into Eq.~(\ref{eq:fractional_frequency_signal}) yields
\begin{eqnarray}
	\label{eq:fractional_frequency_signal_n=1}
	\frac{d\nu}{\nu_0}(t) &=& \frac{2 \alpha}{c} \int_{0}^{t}  \frac{t'-\tau}{\beta} e^{-\frac{t'-\tau}{\beta}} h_{+}(t' - \tau) dt' \nonumber \\
	& = & \frac{2 \alpha }{c \beta}\int_{\frac{-\tau}{\beta}}^{\frac{t-\tau}{\beta}}  u e^{-u} h_{+}(u) du.
\end{eqnarray}
This equation gives a nonzero value for $t > \tau$, and is zero otherwise. Therefore
\begin{eqnarray}
	\label{eq:fractional_frequency_signal_n=1_part2}
	\frac{d\nu}{\nu_0}(t) & = & h_{+}(t - \tau)  \frac{2 \alpha}{c\beta} \int_{0}^{\frac{t-\tau}{\beta}}  u e^{-u} du \nonumber \\
	& = & \frac{2 \alpha}{c\beta}  \left[ - e^{-u} \left( u + 1 \right)\right]_{0}^{\frac{t-\tau}{\beta}} h_{+}(t - \tau)   \nonumber \\
	& = & \frac{2 \alpha}{c\beta} \left( 1  - e^{-\frac{t-\tau}{\beta}} \left( \frac{t-\tau}{\beta} + 1 \right) \right)h_{+}(t - \tau).
\end{eqnarray}
This function is similar to a step function starting from a value zero and rising at
\begin{eqnarray}
	\lim_{t \rightarrow + \infty} \frac{d\nu}{\nu_0}(t) = \frac{2 \alpha}{c\beta}.
\end{eqnarray}
	
\bibliographystyle{unsrt85}
\bibliography{references}
	
\end{document}